%% file: JUNOB8Solar.tex
\documentclass[11pt,a4paper]{article}
\usepackage{graphicx}
\usepackage{url}
\usepackage{amssymb}
\usepackage[centertags]{amsmath}
\usepackage{multirow}
\usepackage{subfigure}
\usepackage[affil-it]{authblk}
\usepackage{cite}
\usepackage{notoccite}
\usepackage{lineno}
\usepackage{tocbibind}
\usepackage[colorlinks,linkcolor=blue,anchorcolor=blue,citecolor=blue,urlcolor=blue]{hyperref}

\newcommand{\nuebar}{$\overline{\nu}_{e}$}
\newcommand{\nue}{$\nu_{e}$}
\newcommand{\numu}{$\nu_{\mu}$}
\newcommand{\nutau}{$\nu_{\tau}$}
\newcommand{\B}{$^8$B}
\newcommand{\T}{$\sin^2\theta_{12}$}
\newcommand{\dm}{$\Delta m^2_{21}$}
\newcommand{\U}{$^{238}$U}
\newcommand{\Th}{$^{232}$Th}
\newcommand{\BiH}{$^{214}$Bi}
\newcommand{\BiL}{$^{212}$Bi}
\newcommand{\PoH}{$^{214}$Po}
\newcommand{\PoL}{$^{212}$Po}
\newcommand{\Tl}{$^{208}$Tl}
\newcommand{\Pa}{$^{234}$Pa$^{\rm m}$}
\newcommand{\Ac}{$^{228}$Ac}
\newcommand{\minitab}[2][l]{\begin{tabular}{#1}#2\end{tabular}}

\begin{document}

\title{Feasibility and physics potential of detecting \B~solar neutrinos at JUNO}
\input{author.tex}
\date{\today}
\maketitle
\begin{abstract}
\noindent
The Jiangmen Underground Neutrino Observatory~(JUNO) features a 20~kt multi-purpose underground liquid scintillator sphere as its main detector.
Some of JUNO's features make it an excellent experiment for \B~solar neutrino measurements, such as its low-energy threshold, its high energy resolution compared to water Cherenkov detectors, and its much large target mass compared to previous liquid scintillator detectors.
In this paper we present a comprehensive assessment of JUNO's potential for detecting \B~solar neutrinos via the neutrino-electron elastic scattering process.
A reduced 2~MeV threshold on the recoil electron energy is found to be achievable assuming the intrinsic radioactive background \U~and \Th~in the liquid scintillator can be controlled to 10$^{-17}$~g/g.
With ten years of data taking, about 60,000 signal and 30,000 background events are expected.
This large sample will enable an examination of the distortion of the recoil electron spectrum that is dominated by the neutrino flavor transformation in the dense solar matter, which will shed new light on the tension between the measured electron spectra and the predictions of the standard three-flavor neutrino oscillation framework.
If $\Delta m^{2}_{21}=4.8\times10^{-5}~(7.5\times10^{-5})$~eV$^{2}$, JUNO can provide evidence of neutrino oscillation in the Earth at the about 3$\sigma$~(2$\sigma$) level by measuring the non-zero signal rate variation with respect to the solar zenith angle.
Moveover, JUNO can simultaneously measure $\Delta m^2_{21}$ using \B~solar neutrinos to a precision of 20\% or better depending on the central value and to sub-percent precision using reactor antineutrinos.
A comparison of these two measurements from the same detector will help elucidate the current tension between the value of $\Delta m^2_{21}$ reported by solar neutrino experiments and the KamLAND experiment.


\end{abstract}


\section{Introduction}
%
Solar neutrinos, produced during the nuclear fusion in the solar core, have played an important role in the history of neutrino physics, from the first observation and appearance of the solar neutrino problem at the Homestake experiment~\cite{Davis:1968cp}, to the measurements at Kamiokande~\cite{Hirata:1989zj}, GALLEX/GNO~\cite{Anselmann:1992um,Altmann:2000ft}, and SAGE~\cite{Abazov:1991rx},
and then to the precise measurements at Super-Kamiokande~\cite{Fukuda:1998fd}, SNO~\cite{Ahmad:2001an,Ahmad:2002jz}, and Borexino~\cite{Arpesella:2007xf}.
In the earlier radiochemical experiments only the charged-current~(CC) interactions of \nue~on the nuclei target could be measured.
Subsequently, solar neutrinos were detected via the neutrino electron elastic scattering~(ES) process in water Cherenkov or liquid scintillator detectors, which are predominantly sensitive to \nue~with lower cross sections for \numu~and \nutau.
Exceptionally, the heavy water target used by SNO allowed observations of all the three processes, including $\nu-e$ ES, CC, and neutral-current~(NC) interactions on deuterium~\cite{Chen:1985na}.
The NC channel is equally sensitive to all active neutrino flavors allowing a direct measurement of the \B~solar neutrino flux at production.
Thus, SNO gave the first model-independent evidence of the solar neutrino flavor conversion and solved the solar neutrino problem.

At present, there are still several open issues to be addressed in solar neutrino physics.
The solar metallicity problem~\cite{Villante:2013mba,Bergemann2014} will profit from either the precise measurements of the $^7$Be and \B~solar neutrino fluxes, or the observation of solar neutrinos from the CNO cycle.
In the elementary particle side, validation tests of the large mixing angle~(LMA) Mikheyev-Smirnov-Wolfenstein~(MSW)~\cite{Wolfenstein:1978ue,Mikheev:1986gs} solution and the search for new physics beyond the standard scenario~\cite{Maltoni:2015kca} constitute the main goals.
The standard scenario of three neutrino mixing predicts a smooth upturn in the \nue~survival probability~($P_{ee}$) in the neutrino energy region between the high~(MSW dominated) and the low~(vacuum dominated) ranges.
However, neither SNO~\cite{Aharmim:2011vm} nor Super-Kamiokande~\cite{Abe:2016nxk} have observed such a spectral upturn in the ES electron spectra, which poses a mild tension with the standard MSW prediction.
Moreover, the indication of the non-zero Day-Night asymmetry in Super-Kamiokande at the level of -$3.3\%$ is much larger than the MSW prediction using the \dm~value from KamLAND~\cite{Gando:2013nba}.
Therefore, a mild inconsistency at the 2$\sigma$ level for the mass-squared splitting $\Delta m^{2}_{21}$ emerges.
%
The combined Super-K and SNO fitting favors $\Delta m^{2}_{21}=4.8^{+1.3}_{-0.6}\times 10^{-5}$~eV$^{2}$~\cite{Abe:2016nxk}, while the long baseline reactor experiment KamLAND gives $\Delta m^{2}_{21}=7.53^{+0.18}_{-0.18}\times 10^{-5}$~eV$^{2}$~\cite{Gando:2013nba}.

To resolve whether this tension is a statistical fluctuation or a physical effect beyond the standard neutrino oscillation framework, requires further measurements.
The Jiangmen Underground Neutrino Observatory~(JUNO), a 20~kt multi-purpose underground liquid scintillator detector, can measure \dm~to an unprecedented sub-percent level using reactor antineutrinos~\cite{An:2015jdp}.
Measurements of \B~solar neutrinos will also benefit primarily due to the large target mass, which affords excellent self-shielding and comparable statistics to Super-K.
A preliminary discussion of the radioactivity requirements and the cosmogenic isotope background can be found in the JUNO Yellow Book~\cite{An:2015jdp}.
In this paper we present a more comprehensive study with the following updates.
The cosmogenic isotopes are better suppressed with improved veto strategies.
The analysis threshold can be lowered to 2~MeV assuming an achievable intrinsic radioactivity background level, which compares favourably with the current world-best 3~MeV threshold in Borexino~\cite{Agostini:2017cav}.
The lower threshold leads to larger signal statistics and a more sensitive examination on the spectrum distortion of recoil electrons.
New evidence of non-zero signal rate variation versus the solar zenith angle~(Day-Night asymmetry) is also expected.
After combining with the \B~neutrino flux from the SNO NC measurement, the \dm~precision is expected to be similar to the current global fitting results~\cite{Esteban:2018azc}.
This paper has the following structure: Sec.~\ref{Signal} presents the expected \B~neutrino signals in the JUNO detector.
Section~\ref{Background} describes the background budget, including the internal and external natural radioactivity, and the cosmogenic isotopes.
Section~\ref{Result} summarizes the results of sensitivity studies.

\section{Solar neutrino detection at JUNO}
\label{Signal}
In LS detectors the primary detection channel of solar neutrinos is their elastic scattering with electrons.
The signal spectrum is predicted with the following steps: generation of neutrino flux and energy spectrum considering oscillation in the Sun and the Earth, determination of the recoil electron rate and kinematics, and implementation of the detector response.
A two-dimensional spectrum of signal counts with respect to the visible energy and the solar zenith angle is produced and utilized in sensitivity studies.
\subsection{\B~neutrino generation and oscillation}

\begin{figure}[h]
\centering
\includegraphics[width=0.6\textwidth]{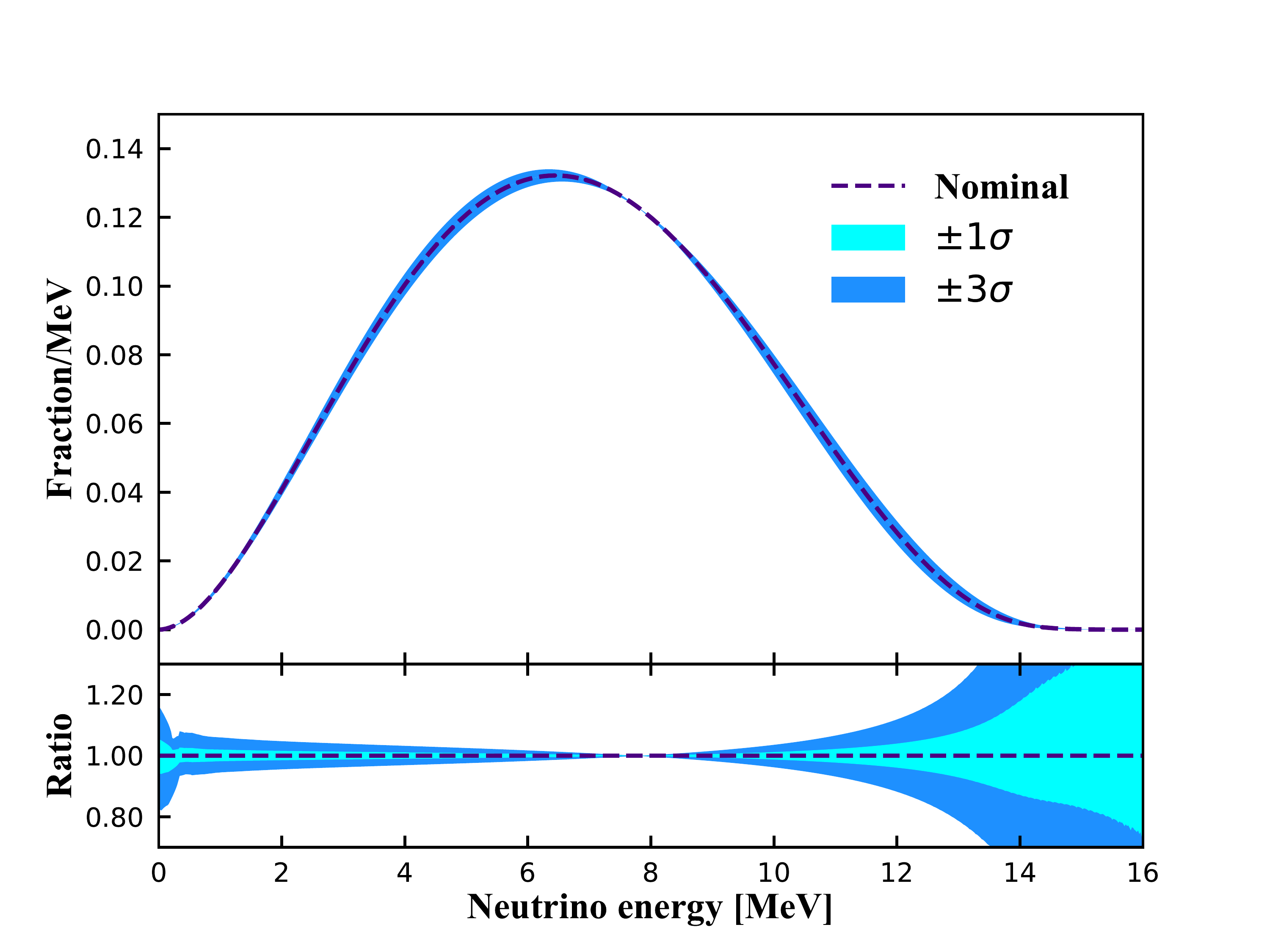}
\caption{\label{fig:B8Shape} \B~\nue~spectrum together with the shape uncertainties. The data are taken from Ref.~\cite{Bahcall:1996qv}.
}
\end{figure}

This study starts with an arrival \B~neutrino flux of (5.25$\pm0.20)\times10^6$~/cm$^2$/s provided by the NC channel measurement at SNO~\cite{Aharmim:2011vm}.
The relatively small contribution~($8.25\times10^3$~/cm$^2$/s) from $hep$~neutrinos, produced by the capture of protons on $^3$He, is also included.
The \B~and $hep$ neutrino spectra are taken from Refs.~\cite{Bahcall:1996qv, Bahcall:1997eg} as shown in Fig.~\ref{fig:B8Shape}.
The neutrino spectrum shape uncertainties~(a shift of about $\pm$100~keV) are mainly due to the uncertain energy levels of $^8$Be excited states.
The shape uncertainties are propagated into the energy-correlated systematic uncertainty on the recoil electron spectrum.
The radial profiles of the solar neutrino production are taken from Ref.~\cite{Bahcall:2004pz}.

The calculation of solar neutrino oscillation in the Sun follows the standard MSW framework~\cite{Wolfenstein:1978ue,Mikheev:1986gs}.
The oscillation is affected by the coherent interactions with the medium via forward elastic weak CC scattering in the Sun.
The neutrino evolution function can be modified by the electron density in the core, which is the so-called MSW effect.
However, due to the slow change in electron density, the neutrino evolution function reduces to an adiabatic oscillation from the position where the neutrino is produced to the surface of the Sun.
Moreover, effects of evolution phases with respect to the effective mass eigenstates average out to be zero due to the huge propagation distance of the solar neutrinos,
resulting in decoherent mass eigenstates prior to arrival at Earth.
The survival probability of solar neutrinos is derived by taking all these effects into account.

\begin{figure}[h]
\centering
\includegraphics[width=0.6\textwidth]{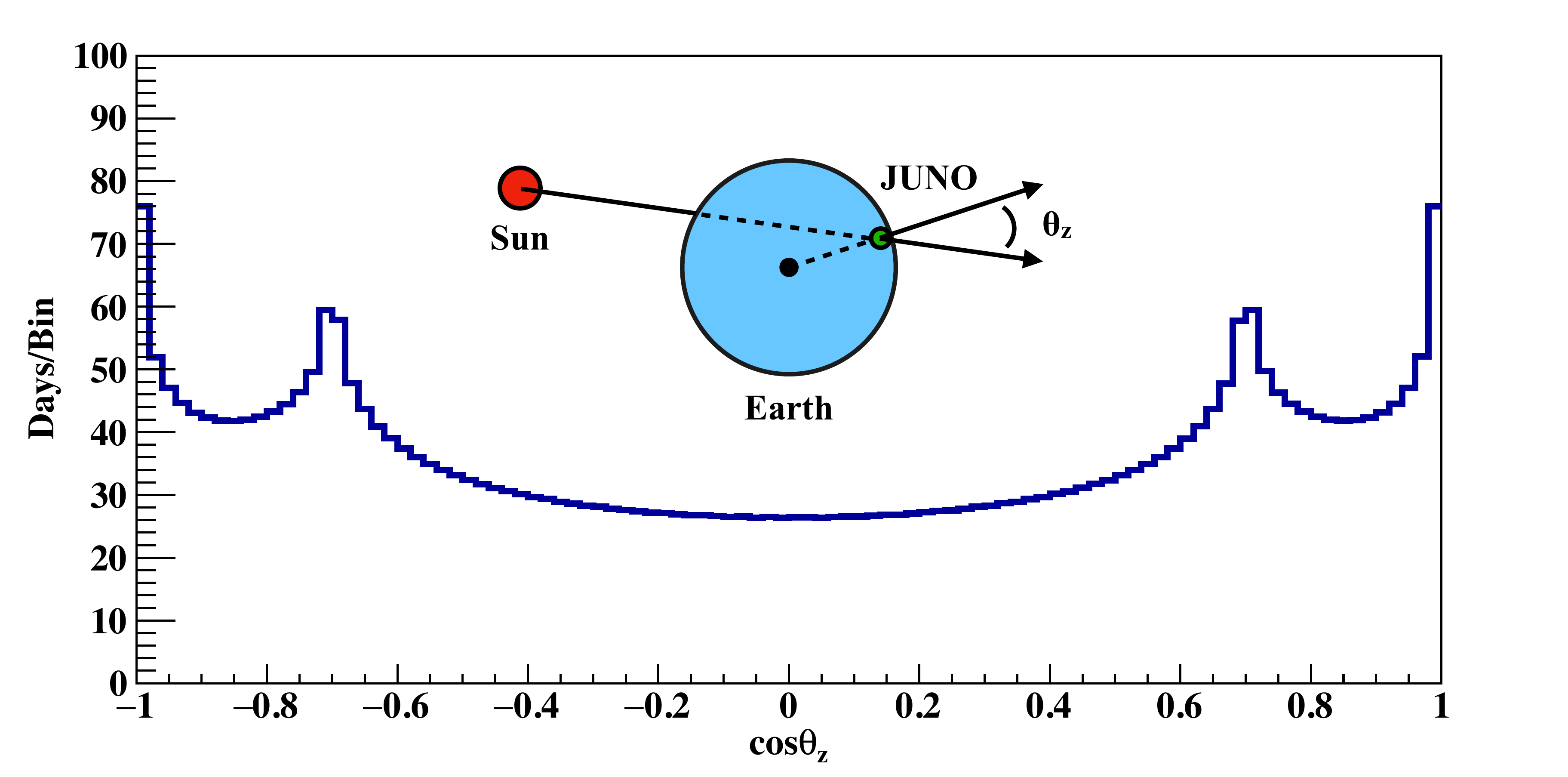}
\caption{\label{fig:zenithangle} Definition of the solar zenith angle $\theta_z$, and the effective detector exposure with respect to 1 A.U.~in ten years of data taking.
}
\end{figure}

During the Night solar neutrinos must pass through the Earth prior to reaching the detector, which via the MSW effect
can make the effective mass eigenstates coherent again, leading to \nue~regenerations.
Compared to Super-K~(36$^{\rm o}$N), the lower latitude of JUNO~(22$^{\rm o}$N) slightly enhances this regeneration.
This phenomenon is quantized by measuring the signal rate variation versus the cosine of the solar zenith angle~($\cos\theta_z$).
The definition of $\theta_z$ and the effective detector exposure with respect to 1~A.U.~in 10 years of data taking are shown in Fig.~\ref{fig:zenithangle}.
In the exposure calculation, the sub-solar points are calculated with the python library PyEphem~\cite{PyEphem} and the Sun-Earth distances given by the library AACGM-v2~\cite{aacgmv2}.
The results are consistent with those in Ref.~\cite{Ioannisian:2015qwa}.
The Day is defined as $\cos\theta_z<0$, and the Night as $\cos\theta_z>0$.
The \nue~regeneration probability is calculated assuming a spherical Earth and using the averaged 8-layer density from the Preliminary Reference Earth Model~(PREM)~\cite{Dziewonski:1981xy}.

\begin{figure}[!htb]
\centering
\includegraphics[width=0.6\textwidth]{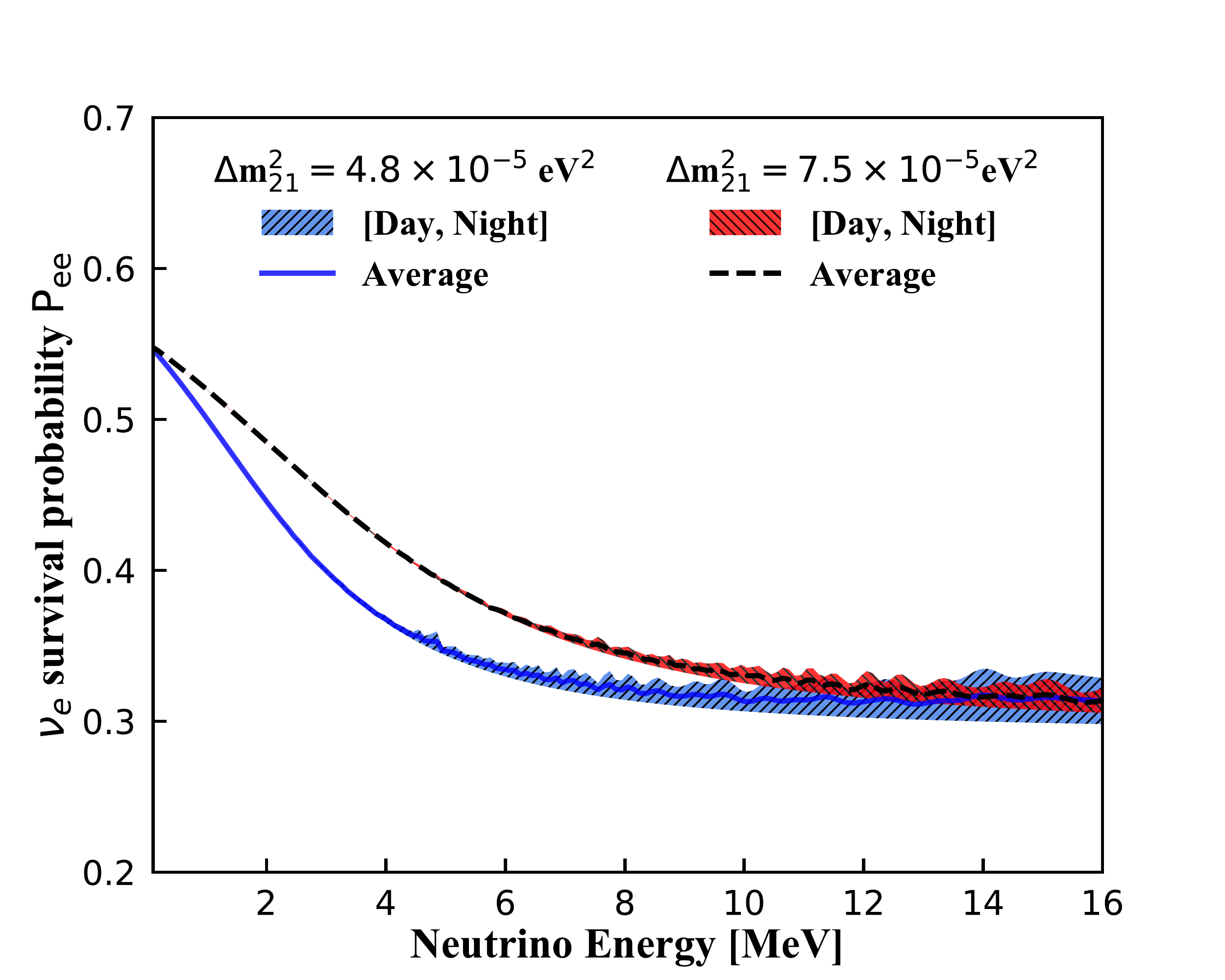}
\caption{\label{fig:Pee} Solar \nue~survival probabilities~($P_{ee}$) with respect to the neutrino energy. A transition from the MSW dominated oscillation to the vacuum dominated is found when neutrino energy goes from high to low ranges. The shadowed area shows the variation of $P_{ee}$ at different solar zenith angles. A smaller \dm~leads to a larger Day-Night asymmetry effect.
}
\end{figure}

Taking the MSW effects in both the Sun and the Earth into consideration, the \nue~survival probabilities~($P_{ee}$) with respect to the neutrino energy $E_\nu$
for two \dm~values are shown in Fig.~\ref{fig:Pee}.
The other oscillation parameters are taken from PDG-2018~\cite{Tanabashi:2018oca}.
The shadowed area shows the $P_{ee}$ variation at different solar zenith angles.
A transition energy range is apparent where matter effects are present but not fully expressed.
A smooth upturn trend of $P_{ee}(E_\nu)$ in this transition energy range is also expected.
The smaller the \dm~value is, the steeper upturn at the transition range and the larger size of Day-Night asymmetry at high energies can be found.
Furthermore, the $P_{ee}(E_\nu)$ in the transition region is especially sensitive to non-standard interactions~\cite{Maltoni:2015kca}.
Thus, by detecting \B~neutrinos, the existence of new physics which sensitively affects the transition region can be tested, and the Day-Night asymmetry can be measured.

\subsection{$\nu-e$ elastic scattering}

\begin{figure}[!htb]
\centering
\includegraphics[width=0.6\textwidth]{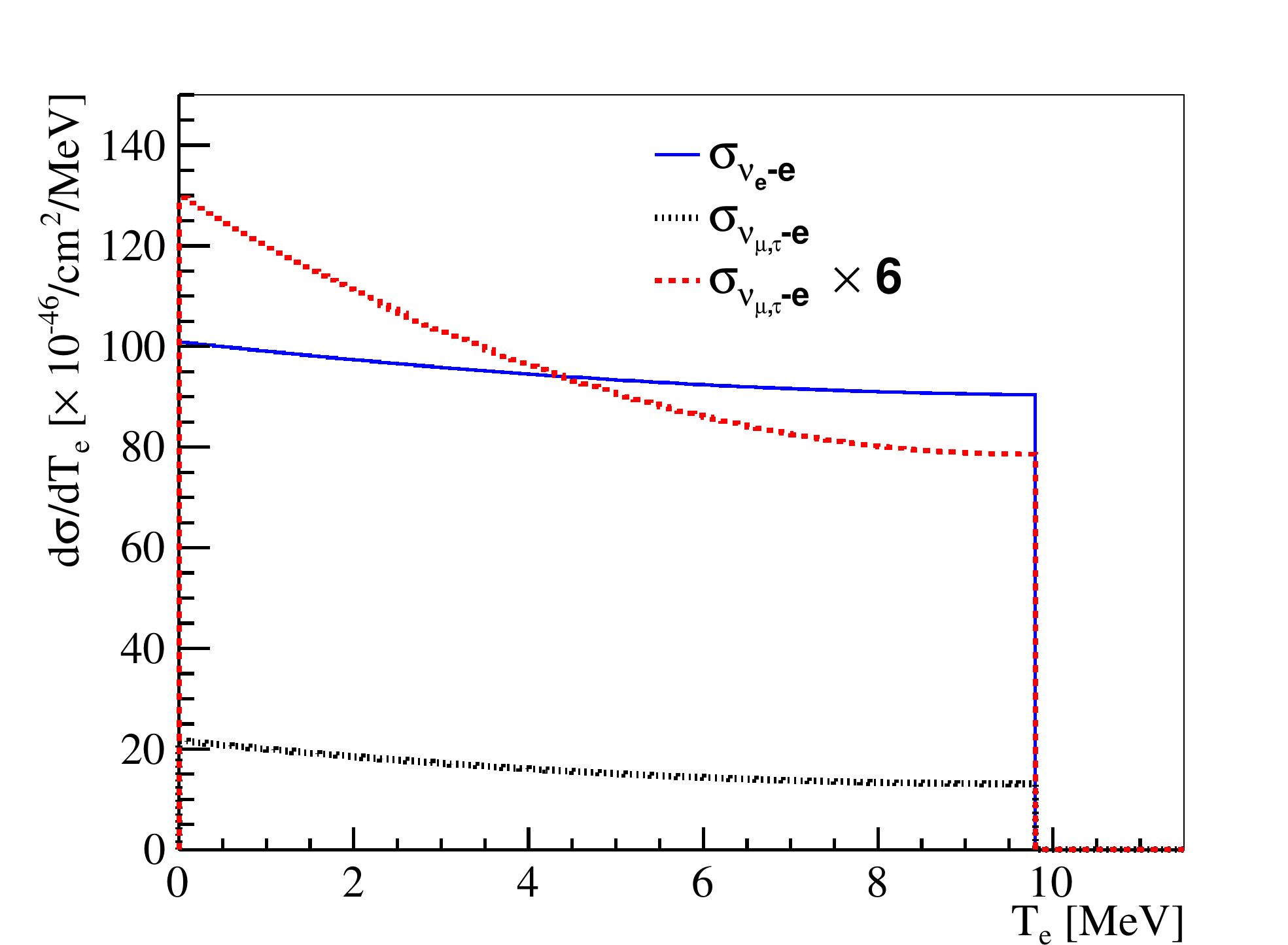}
\caption{\label{fig:Xsection} Differential cross section of $\nu_e-e$~(blue) and $\nu_{\mu,\tau}-e$~(black) elastic scattering for a 10~MeV neutrino. The stronger energy dependence of the $\nu_{\mu,\tau}-e$ cross section, as illustrated in red, produces another smooth upturn in the visible electron spectrum compared to the case of no $\nu_{\mu,\tau}$ appearance.
}
\end{figure}

In the $\nu-e$ elastic scattering process, \nue~can interact with electrons via both $W^\pm$ and $Z^0$ boson exchange, while $\nu_{\mu,\tau}$ can only interact with electrons via $Z^0$ exchange.
This leads to an about six times larger cross section of \nue$-e$) compared to that of $\nu_{\mu,\tau}-e$, as shown in Fig.~\ref{fig:Xsection}.
For the cross section calculation, we use:
\begin{equation}
\frac{d\sigma}{dT_e}(E_{\nu},T_e)=\frac{\sigma_0}{m_e}[g^2_1+g^2_2(1-\frac{T_e}{E_{\nu}})^2-g_1g_2\frac{m_eT_e}{E_{\nu}^2}]~,
\label{equ:crosssection}
\end{equation}
where $E_{\nu}$ is the neutrino energy, $T_e$ is the kinetic energy of the recoil electron, $m_e$ is the electron mass, $\sigma_0=\frac{2G^2_Fm^2_e}{\pi}\simeq88.06\times10^{-46}$ cm$^2$~\cite{Giunti:2007ry}.
The quantities $g_1$ and $g_2$ depend on neutrino flavor: $g_1^{(\nu_e)}=g_2^{(\overline{\nu}_e)}\simeq0.73$, $g_2^{(\nu_e)}=g_1^{(\overline{\nu}_e)}\simeq0.23$, $g_1^{(\nu_{\mu,\tau})}=g_2^{(\overline{\nu}_{\mu,\tau})}\simeq-0.27$, $g_2^{(\nu_{\mu,\tau})}=g_1^{(\overline{\nu}_{\mu,\tau})}\simeq0.23$.

\begin{figure}[!htb]
\centering
\includegraphics[width=0.6\textwidth]{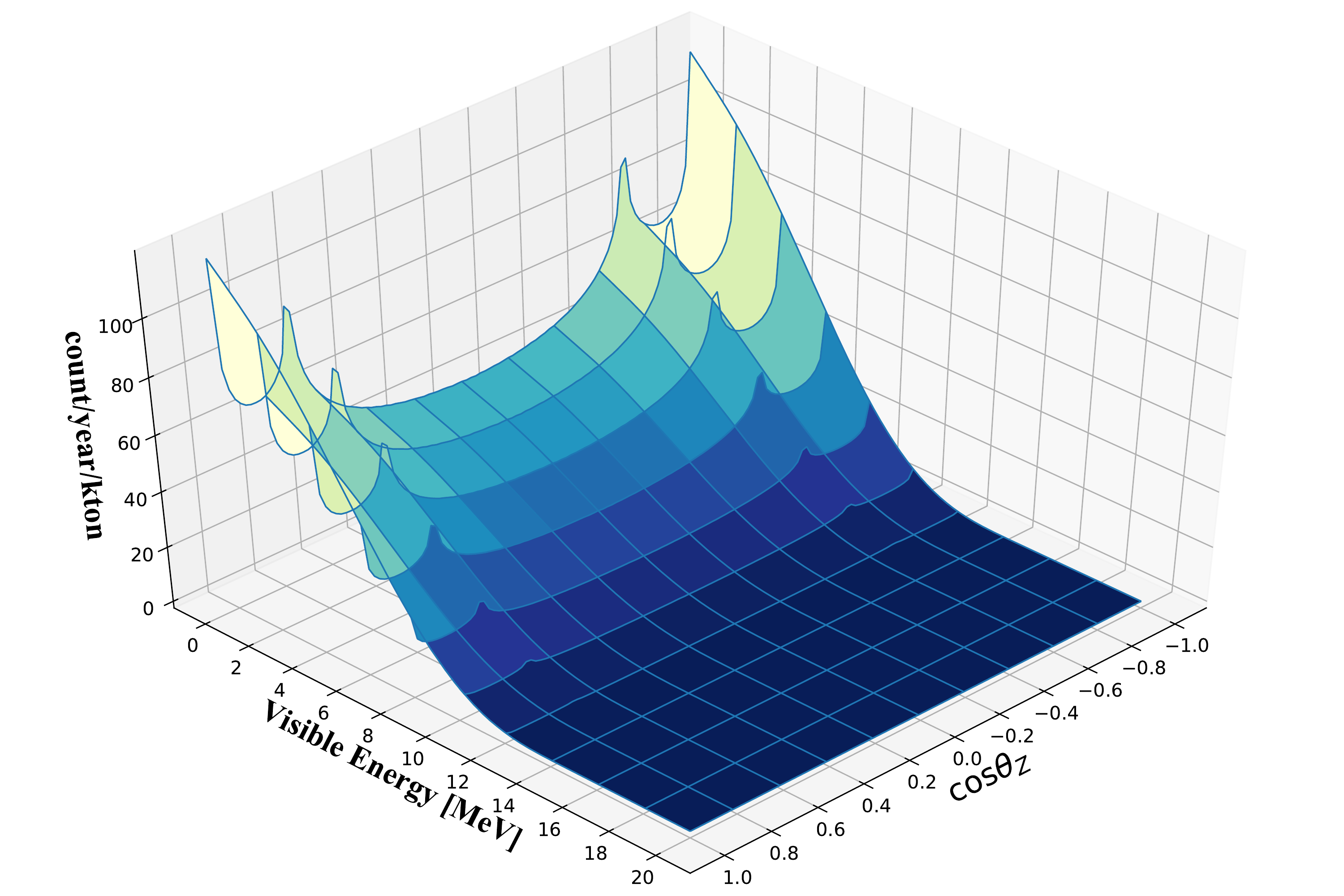}
\caption{\label{fig:2DSignal} \B~$\nu-e$ ES signal counts with respect to the visible energy of recoil electron and the cosine of the solar zenith angle $\cos\theta_z$. The spectrum carries information on neutrino oscillation both in the Sun and the Earth, and will be used to determine the neutrino oscillation parameters.
}
\end{figure}

After scattering, the total energy and momentum of the neutrino and electron are redistributed.
In the JUNO detector the direction of the recoil electron is difficult to reconstruct,
making an event-by-event reconstruction of the neutrino energy almost impossible.
Physics studies rely on the visible spectrum predicted in the following steps.
First, applying the ES cross section to the neutrino spectrum to obtain the kinetic energy spectrum of recoil electrons.
To calculate the reaction rate, the electron density is 3.38$\times10^{32}$ per kt using the LS composition in Ref.~\cite{An:2015jdp}.
The expected signal rate in the full energy range is 4.15~(4.36) counts per day per kt~(cpd/kt) for \dm$=4.8~(7.5)\times 10^{-5}$~eV$^{2}$.
A simplified detector response model, including the LS's light output nonlinearity from Daya Bay~\cite{Adey:2019zfo} and the 3\%/$\sqrt E$ energy resolution, is applied to the kinetic energy of recoil electron, resulting in the visible energy E$_{\rm vis}$.
The ES reaction vertex is also smeared by assuming a 12~cm resolution at 1~MeV.
Eventually, the number of signals are counted with respect to the visible energy and the solar zenith angle as shown in Fig.~\ref{fig:2DSignal}.
The two-dimensional spectrum will be used to determine the neutrino oscillation parameters, since it carries information on both the spectrum distortion primarily from oscillation in the Sun, and the Day-Night asymmetry from oscillation in the Earth.
Table~\ref{table:rate} provides the expected signal rates during Day and Night within two visible energy ranges.

\begin{table}[h]
\begin{center}
	 \begin{tabular}{ccccc}
        \hline
        \multirow{2}{*}{Rate~[cpd/kt]}& \multicolumn{2}{c}{(0,~16)~MeV}&\multicolumn{2}{c}{ (2,~16)~MeV}\\
        \cline{2-5}
        &Day &Night  &Day &Night \\ \hline
        $\Delta m^2_{21}=4.8\times10^{-5}$eV$^2$  & 2.05 & 2.10  & 1.36 & 1.40  \\ \hline
        $\Delta m^2_{21}=7.5\times10^{-5}$eV$^2$  & 2.17 & 2.19  & 1.44 & 1.46  \\
        \hline
	\end{tabular}
	\caption{\label{table:rate} JUNO \B~$\nu-e$ E signal rates in terms of per day per kt~(cpd/kt) during the Day and Night, in different visible energy ranges.}
\end{center}
\end{table}

\section{Background budget}
\label{Background}

Unlike the correlated signals produced in the Inverse Beta Decay reaction of reactor antineutrinos, the ES signal of solar neutrinos is a single event.
A good signal-to-background ratio requires extremely radiopure detector materials, sufficient shielding from surrounding natural radioactivity and an effective strategy to reduce backgrounds from unstable isotopes produced by cosmic-ray muons passing through the detector.
Based on the R\&D of JUNO detector components, a background budget has been built for \B~neutrino detection at JUNO.
Assuming an intrinsic \U~and \Th~radioactivity level of 10$^{-17}$~g/g, the 2~MeV analysis threshold could be achieved, yielding a sample from 10 years of data taking of about 60,000 ES signal events and 30,000 background candidates.

The threshold cannot be further reduced below 2 MeV due to the large background from cosmogenic $^{11}$C,
which is a $\beta^+$ isotope with a decay energy of 1.982~MeV and a half-life of 20.4~minutes
with production rate in the JUNO detector of more than 10,000 per day.
The huge yield and long life-time of $^{11}$C makes it very difficult to suppress this background
to a level similar to the signal, limiting the analysis threshold to 2~MeV.

\begin{figure}[!htb]
\begin{center}
	\includegraphics[width=0.7\textwidth]{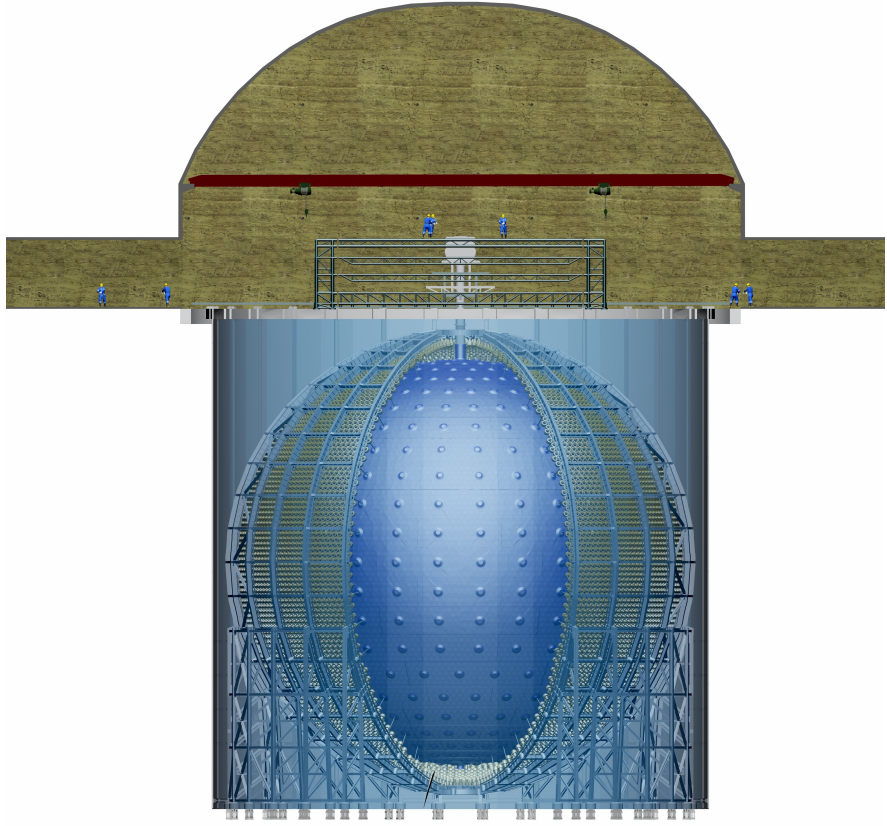}
	\caption{Diagram of the JUNO detector. The 20~kt LS is contained in a spherical acrylic vessel with an inner diameter of 35.4~m, and the vessel is supported by stainless steel latticed shell and the shell also holds about 18,000~pieces of 20-inch PMTs and 25,000 pieces of 3-inch PMTs.}
	\label{fig:JUNODet}
\end{center}
\end{figure}

\subsection{Natural radioactivity}

As shown in Fig.~\ref{fig:JUNODet}, the 20~kt liquid scintillator is contained in a spherical acrylic vessel with an inner diameter of 35.4~m and a thickness of 12~cm.
The vessel is supported by a 600~t stainless steel~(SS) structure composed of 590 SS bars connected to acrylic nodes.
Each acrylic node includes an about 40~kg SS ring providing enough strength.
The LS is instrumented with about 18,000 20-inch PMTs and 25,000 3-inch small PMTs.
The 18,000 20-inch PMTs comprise 5,000 Hamamatsu dynode PMTs and 13,000 PMTs with a microchannel plate~(MCP-PMT) instead of a dynode structure.
All the PMTs are installed on the SS structure and the glass bulbs of the large PMTs are positioned about 1.7~m away from the LS.
Pure water in the pool serves as both passive shielding and a Cherenkov muon detector instrumented with about 2,000 20-inch PMTs.
The natural radioactivity is divided into internal and external parts, where the internal part is the LS intrinsic background
and the external part is from other detector components and surrounding rocks.
The radioactivity of each to-be-built detector component has been measured~\cite{sisti_monica_2018_1300598} and is used in the Geant4~(10.2) based simulation~\cite{Geant4}.
\subsubsection{External radioactivity}

Among the external radioactive isotopes, \Tl~is the most critical one due to that it has the highest energy $\gamma$~(2.61~MeV) from its decay.
This is also the primary reason that Borexino could not lower the analysis threshold to below 3~MeV~\cite{Agostini:2017cav}.
The problem is overcome in JUNO due to its much larger detector size.
In addition, all detector materials of JUNO have been carefully selected to fulfill radiopurity requirements~\cite{Li_2016}.
The \U~and \Th~contaminations in SS are measured to be less than 1~ppb and 2~ppb, respectively.
In acrylic both are at 1~ppt level.
An improvement is from the glass bulbs of MCP-PMTs, in which the \U~and \Th~contaminations are 200~ppb and 125~ppb, respectively~\cite{Zhang:2017ocm}.
Both are several times lower than those of the 20-inch Hamamatsu PMTs.

\begin{figure}[htb]
\begin{center}
	\includegraphics[width=0.6\textwidth]{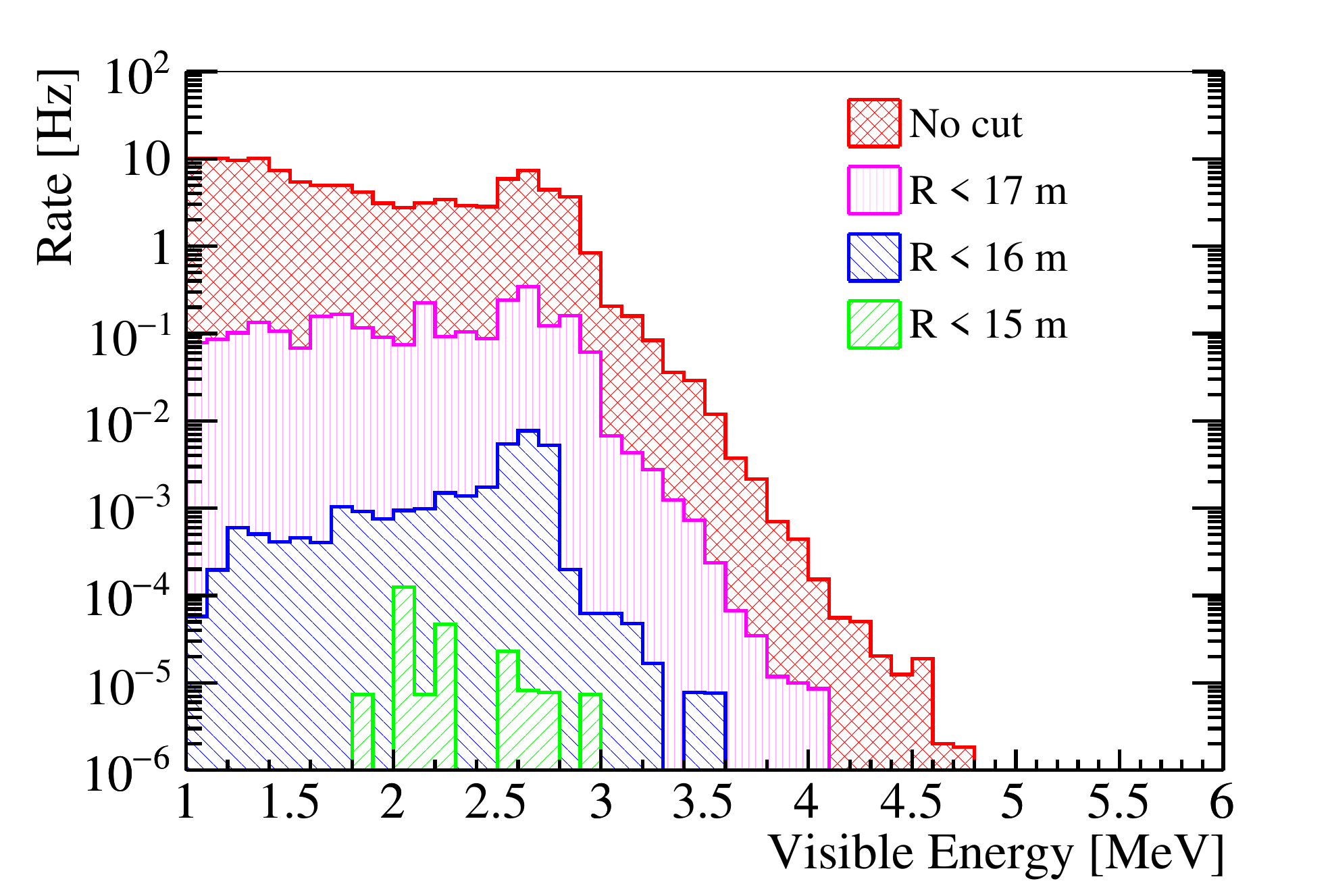}
	\caption{Deposited energy spectra in LS by external $\gamma$'s after different FV cuts. The plot is generated by the simulation with measured radioactivity values of all external components. The multiple $\gamma$'s of \Tl~decays in acrylic and SS nodes account for the background in 3 to 5~MeV. With a set of energy-dependent FV cuts, the external background is suppressed to less than 0.5\% compared to signals.  }\label{fig:ExternalRadio}
\end{center}
\end{figure}

With the measured radioactivity values, a simulation is performed to obtain the external $\gamma$'s deposited energy spectrum in LS as shown in Fig.~\ref{fig:ExternalRadio}.
%
\Tl~ decays in the PMT glass and in the SS rings of the acrylic nodes dominate the external background, which is responsible for the peak at 2.6~MeV.
The continuous part from 2.6 to 3.5~MeV is due to the multiple $\gamma$'s released from a \Tl~decay.
Limited by the huge computing resources required to simulate enough \Tl~decays in the PMT glass, an extrapolation method is used to estimate its contribution in the region of spherical radius less than 15~m based on the simulated results in the outer region.
According to the simulation results, an energy dependent fiducial volume~(FV) cut, in terms of the reconstructed radial position~($r$) in the spherical coordinate system, is designed as:
\begin{itemize}
\item 2$<E_{\rm vis}\le$3~MeV, $r<$13~m, 7.9~kt target mass;
\item 3$<E_{\rm vis}\le$5~MeV, $r<$15~m, 12.2~kt target mass;
\item $E_{\rm vis} >$5~MeV, $r<$16.5~m, 16.2~kt target mass.
\end{itemize}
In this way the external radioactivity background is suppressed to less than 0.5\% compared to signals in the whole energy range, while the signal statistic is maximized at high energies.

In addition to the decays of natural radioactive isotopes, another important source of high energy $\gamma$'s is the ($n,~\gamma$) reaction in rock, PMT glass, and the SS structure~\cite{Agostini:2017cav,Abe:2011em}.
Neutrons mainly come from the ($\alpha,~n$) reaction and the spontaneous fission of \U, and are named as radiogenic neutrons.
The neutron fluxes and spectra are calculated using the neutron yields from Refs.~\cite{Agostini:2017cav,Zhao:2013mba} and the measured radioactivity levels.
Then the simulation is performed to account for the neutron transportation and capture, high energy $\gamma$ release and energy deposit.
In the FV of $r<$16.5~m, this radiogenic neutron background contribution is found to be less than 0.001 per day and can be neglected.

\subsubsection{Internal radioactivity}

\begin{figure}[htb]
\begin{center}
	\includegraphics[width=0.4\textwidth]{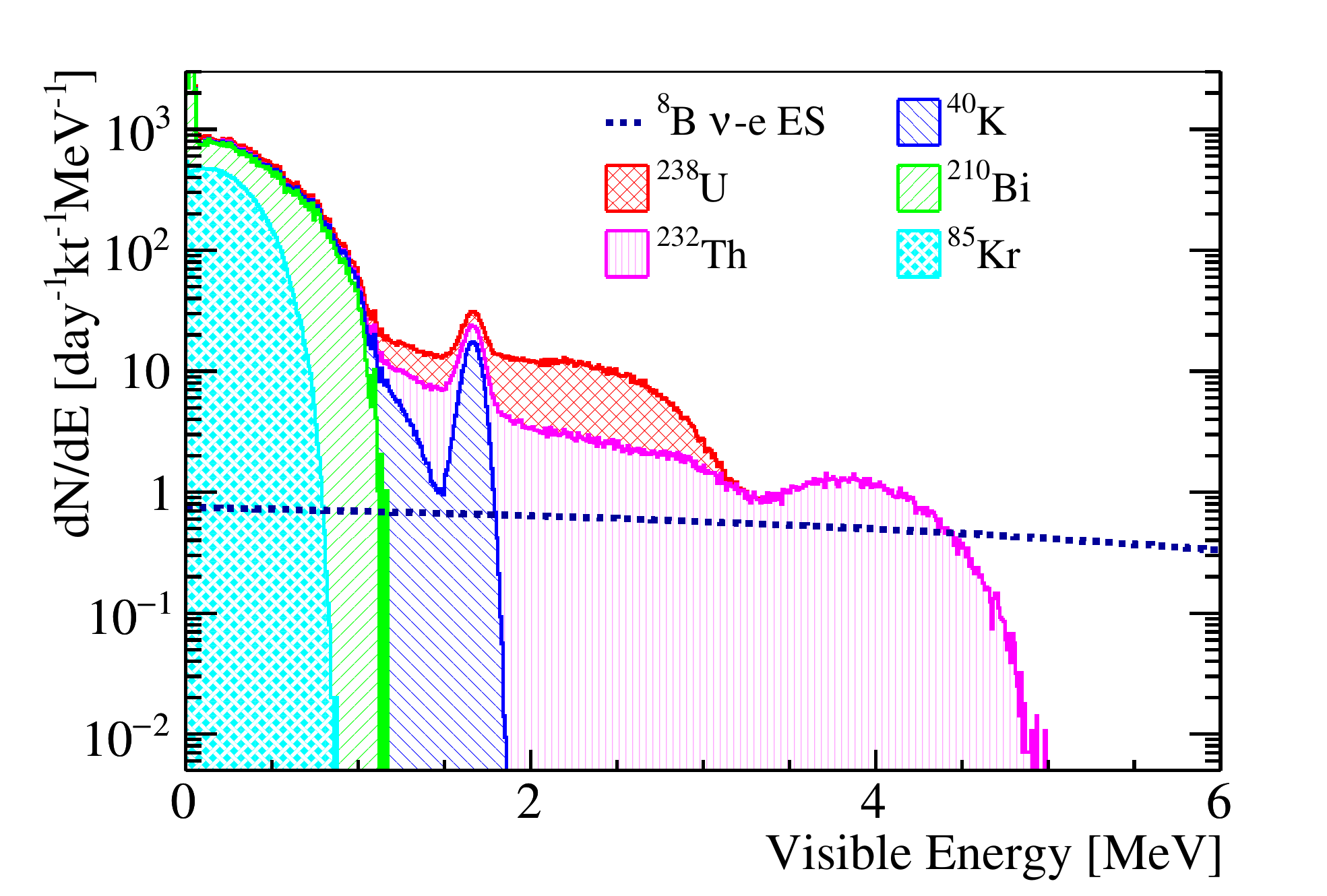}
	\includegraphics[width=0.4\textwidth]{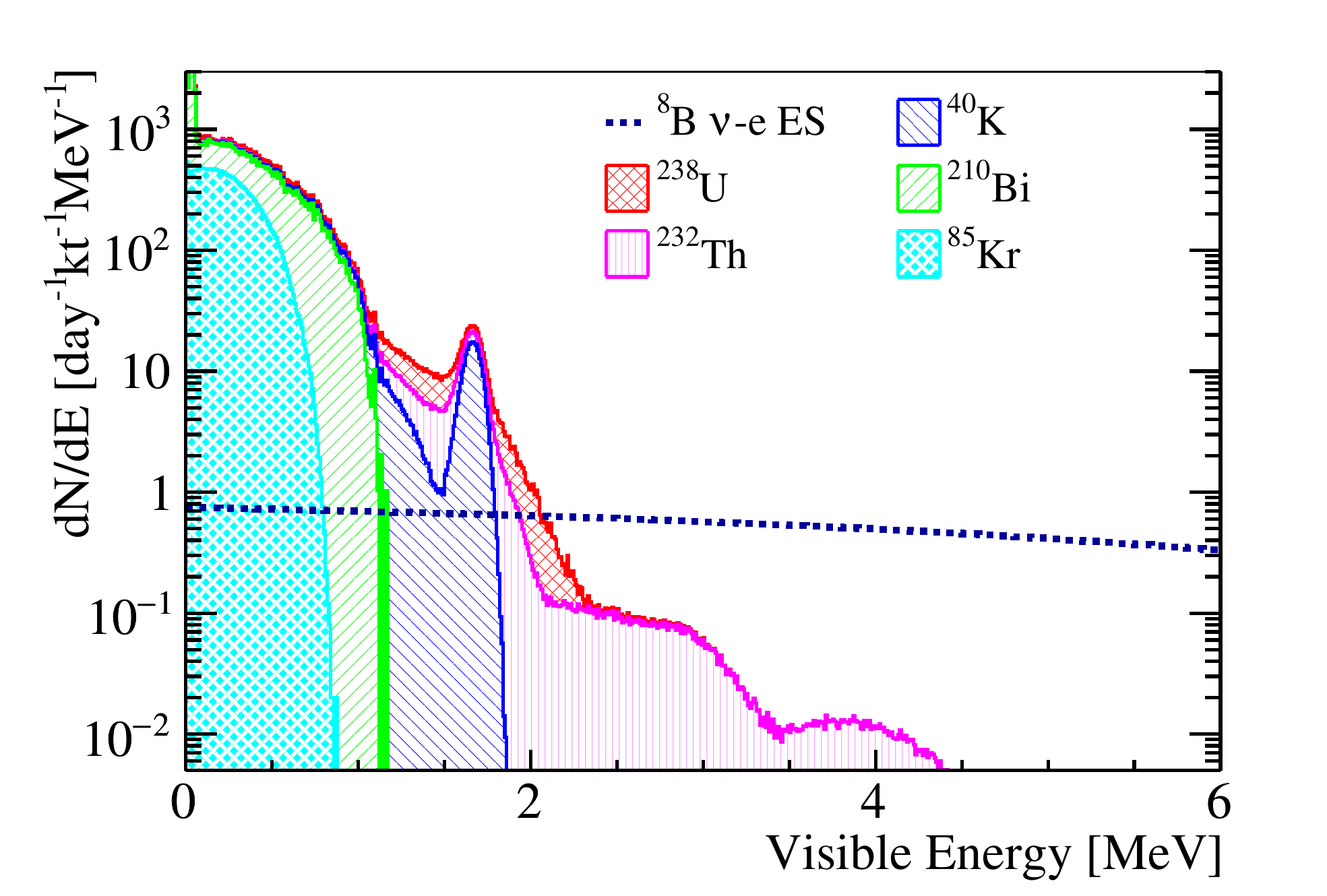}
	\caption{Internal radioactivity background compared with \B~signal before~(left) and after~(right) time, space and energy correlation cuts to remove the Bi-Po/Bi-Tl cascade decays. The events in the $3-5$~MeV energy range are dominated by \Tl~decays, while those between 2 and 3~MeV are from \BiH~and \BiL.}\label{fig:Internal}
\end{center}
\end{figure}

With negligible external background after the FV cuts, the LS intrinsic impurity levels and backgrounds from cosmogenic isotopes determine the lower analysis threshold of recoil electrons.
JUNO will deploy four LS purification approaches. Three of them focus on the removal of natural radioactivity in LS during distillation, water extraction and gas stripping~\cite{Distillation}.
An additional online monitoring system~(OSIRIS~\cite{OSIRIS}) will be built to measure the \U, \Th, and Radon contaminations before the LS filling.
As a feasibility study, following the assumptions in the JUNO Yellow Book~\cite{An:2015jdp}, we start with 10$^{-17}$~g/g \U~and \Th~in the secular equilibrium, which are close to those of Borexino Phase I~\cite{Arpesella:2008mt}.
However, from Borexino's measurements the daughter nuclei of $^{222}$Rn, i.e., $^{210}$Pb and $^{210}$Po, are likely off-equilibrium.
In this study $10^{-24}$~g/g $^{210}$Pb and a $^{210}$Po decay rate of 2600~cpd/kt are assumed~\cite{Agostini:2017ixy}.
The left plot of Fig.~\ref{fig:Internal} shows the internal background spectrum under the assumptions above, and the \B~neutrino signal is also drawn for comparison.
Obviously, an effective background reduction method is needed.
The $\alpha$ peaks are not included since after LS quenching their visible energies are usually less than 1~MeV, much smaller than the 2~MeV analysis threshold.

Above the threshold the background is dominated by five isotopes as listed in Table~\ref{tab:radioInternal}.
\BiH~ and 64\% of \BiL~decays can be removed by the coincidence with their short-lived daughter nuclei, \PoH~($\tau\sim231~\mu$s)~and \PoL~($\tau\sim431$~ns), respectively.
The removal efficiency of $^{214}$Bi can reach to about 99.5\% with less than 1\% loss of signals.
Based on the current electronics design, which records PMT waveforms in a 1~$\mu$s readout window with a sampling rate of 1~G samples/s~\cite{Bellato:2020lio}, it is assumed that \PoL~cannot be identified from its parent \BiL~if it decays within 30~ns.
Thus, for the \BiL-\PoL~cascade decays the removal efficiency is only 93\%.
For the residual 7\%, the visible energies of \BiL~and \PoL~decays are added together because they are too close to each other.

Besides removing \BiH~and \BiL~via the prompt correlation which was also used in previous experiments, a new analysis technique of this study is the reduction of \Tl.
36\% of \BiL~decays to \Tl~by releasing an $\alpha$ particle.
The decay of \Tl~($\tau\sim$4.4~minutes) dominates the background in the energy range of 3 to 5~MeV.
With a 22~minutes veto in a spherical volume of radius 1.1~m around a \BiL~$\alpha$ candidate, 99\% \Tl~decays can be removed.
The fraction of removed good events, estimated with the simulation, is found to be about 20\%, because there are more than 2600~cpd/kt $^{210}$Po decays in the similar $\alpha$ energy range.
Eventually, the signal over background ratio in the energy range of $3-5$~MeV is significantly improved from 0.6 to 35.

\begin{table*}[htb]
\small
\begin{center}
	\begin{tabular}{c|c|c|c|c|c|c|c}
        \hline
        Isotope   & Decay mode & Decay energy & $\tau$ & Daughter & Daughter's $\tau$ & Removal eff.  & Removed signal\\ \hline
        \BiH & $\beta^-$       & 3.27~MeV & 28.7~min & \PoH      & 237~$\mu$s &$>$99.5\% & $<$1\% \\
        \BiL & $\beta^-$: 64\% & 2.25~MeV & 87.4~min & \PoL      & 431~ns & 93\% & $\sim$0 \\
		\BiL & $\alpha$: 36\%  & 6.21~MeV & 87.4~min & \Tl       & 4.4~min&  N/A & N/A \\
		\Tl  & $\beta^-$       & 5.00~MeV & 4.4~min  & $^{208}$Pb& Stable & 99\%  & 20\% \\
		\Pa  & $\beta^-$       & 2.27~MeV & 1.7~min  & $^{234}$U & 245500 years & N/A  & N/A \\
		\Ac  & $\beta^-$       & 2.13~MeV & 8.9~h    & $^{228}$Th& 1.9 years & N/A  & N/A \\ \hline
	\end{tabular}
	\caption{Isotopes in the \U~and \Th~decay chains with decay energies larger than 2~MeV. With correlation cuts most of \BiH, \BiL~and \Tl~decays can be removed. The decay data are taken from Ref.~\cite{ATOM}.}
	\label{tab:radioInternal}
\end{center}
\end{table*}

However, for \Ac~and \Pa, both of which have decay energies slightly larger than 2~MeV, there are no available cascade decays for background elimination.
If the \U~and \Th~decay chains are in secular equilibrium, their contributions can be statistically subtracted with the measured Bi-Po decay rates.
Otherwise the analysis threshold should be increased to about 2.3~MeV.

Considering higher radioactivity level assumptions, if the \U~and \Th~contaminations are $10^{-16}$~g/g, the 2~MeV threshold is still achievable but with a worse S/B ratio in the energy range of 3 to 5~MeV.
A $10^{-15}$~g/g contamination would result in a 5~MeV analysis threshold determined by the end-point energy of \Tl~ decay.
If \U~and \Th~contaminations reach to $10^{-17}$~g/g, but the $^{210}$Po decay rate is more than 10,000~cpt/kt as in the Borexino Phase~I~\cite{Arpesella:2008mt}, the \Tl~reduction mentioned above cannot be performed.
Consequently, the \Tl~background could only be statistically subtracted.
The influence of these radioactivity level assumptions on the neutrino oscillation studies will be discussed in Sec.~\ref{OscillationResults}.

\subsection{Cosmogenic isotopes}

\begin{figure}[htb]
\begin{center}
	\includegraphics[width=0.4\textwidth]{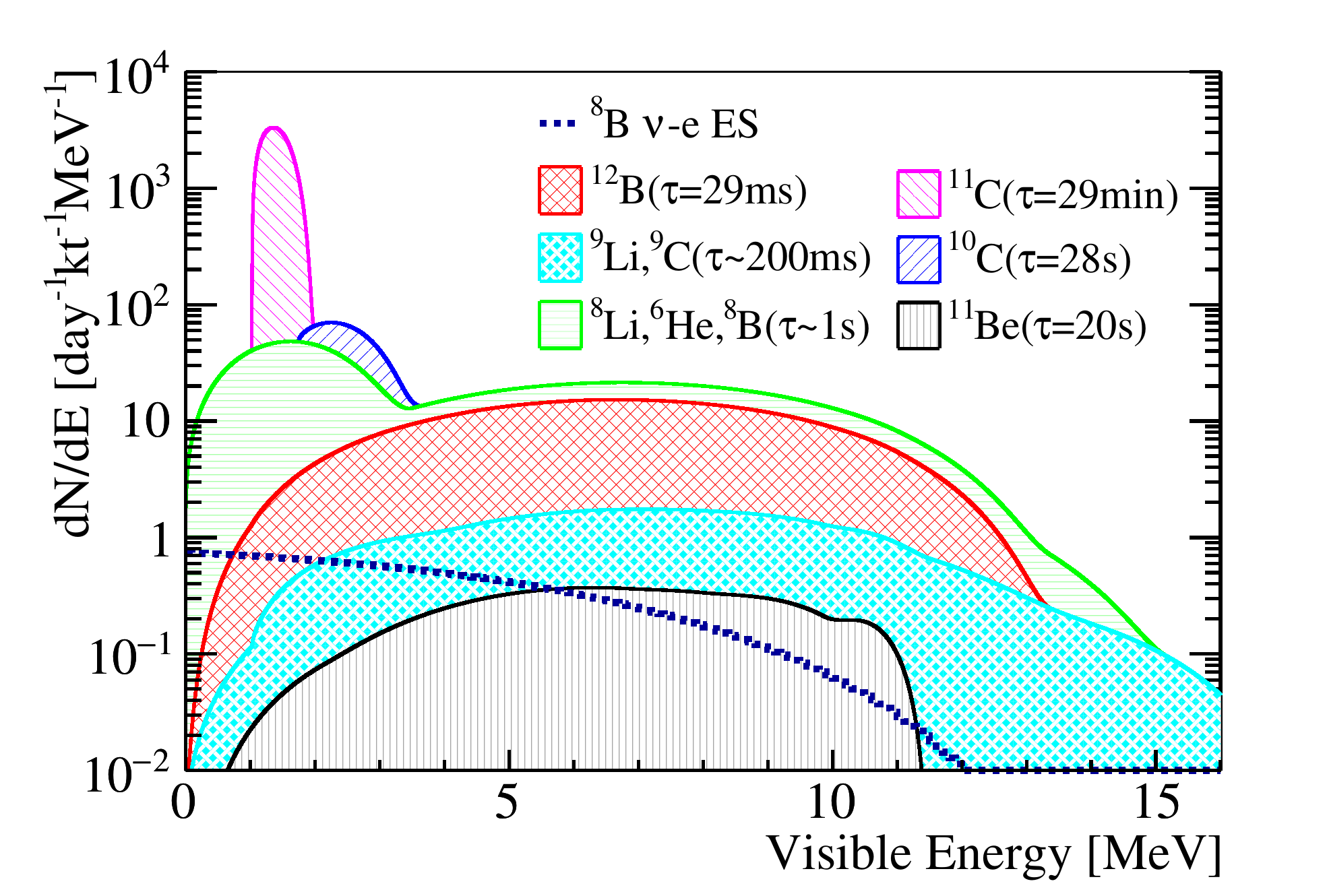}
	\includegraphics[width=0.4\textwidth]{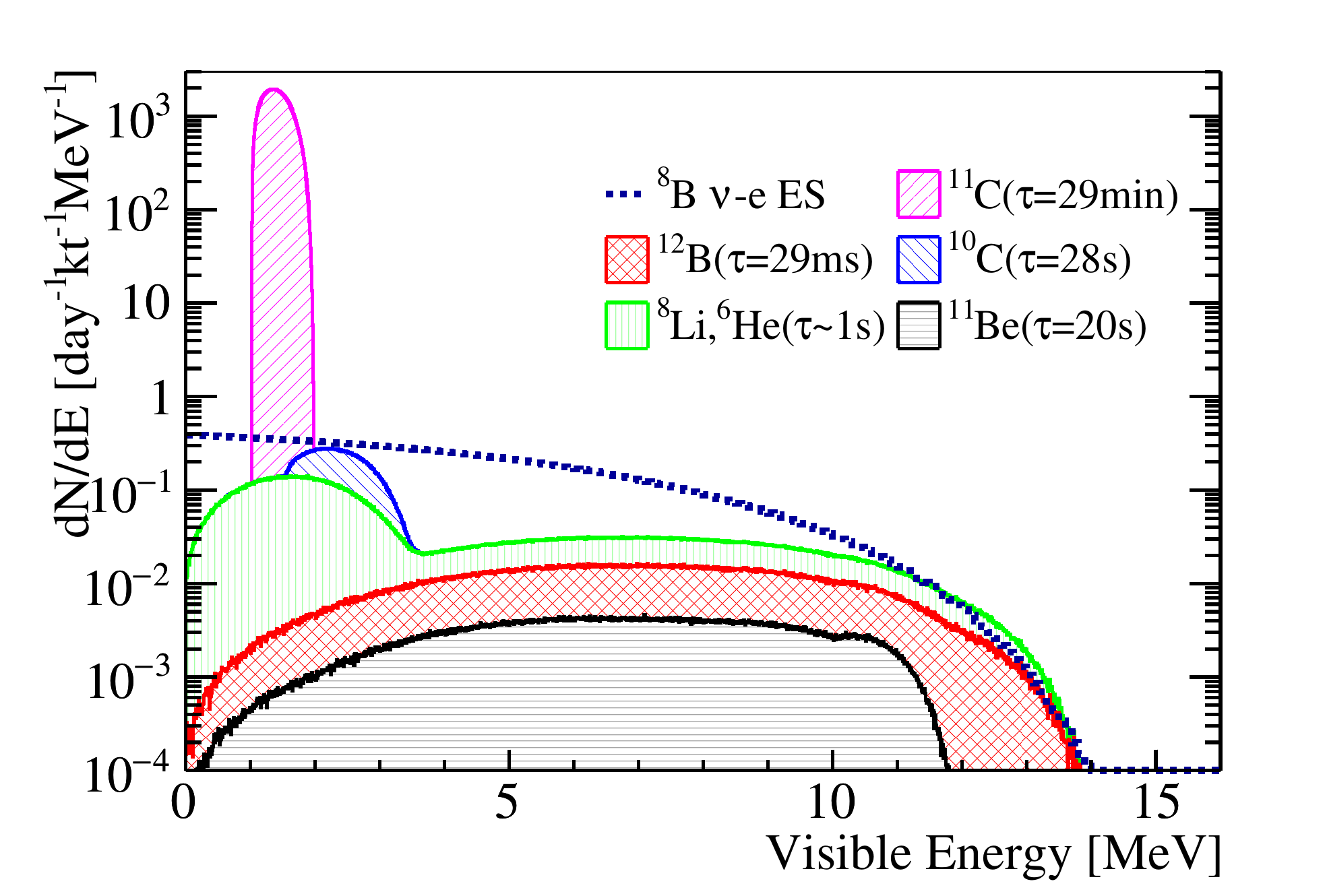}
	\caption{ Cosmogenic background before~(left) and after veto~(right). The isotope yields shown here are scaled from the KamLAND's~\cite{Abe:2009aa} and Borexino's measurements~\cite{Bellini:2008mr}. The huge amount of $^{11}$C constrains the analysis threshold to 2~MeV. The others isotopes can be well suppressed with veto strategies discussed in the text. }\label{fig:CosmogenicIsotope}
\end{center}
\end{figure}

In addition to natural radioactivity, another crucial background comes from the decays of light isotopes produced by the cosmic-ray muon spallation process in LS.
The relatively shallow vertical rock overburden, about 680~m, leads to a 0.0037~Hz/m$^2$ muon flux with an averaged energy of 209~GeV.
The direct consequence is the about 3.6~Hz muons passing through the LS target.
More than 10,000 $^{11}$C isotopes are generated per day, which constrains the analysis threshold to 2~MeV, as shown in Fig.~\ref{fig:CosmogenicIsotope}.
Based on the simulation and measurements of previous experiments, it is found that other isotopes can be suppressed to a 1\% level with a cylindrical veto along the muon track and the Three-Fold Coincidence cut~(TFC) among the muon, the spallation neutron capture, and the isotope decay~\cite{Balata:2006ue,Agostini:2018uly}.
Details are presented in this section.

\subsubsection{Isotope generation}

When a muon passes through the LS, along with the ionization, many secondary particles are also generated, including $e^\pm$, $\gamma$, $\pi^{\pm}$, and $\pi^{\rm 0}$.
Neutrons and isotopes are produced primarily via the ($\gamma$,~n) and $\pi$~inelastic scattering processes.
More daughters could come from the neutron inelastic scattering on carbon.
Such a process is defined as a hadronic shower in which most of the cosmogenic neutrons and light isotopes are generated.
More discussion on the muon shower process can be found in Refs.~\cite{Li:2015kpa,Li:2015lxa}.
To understand the shower physics and develop a reasonable veto strategy, an detailed muon simulation has been carried out.
The simulation starts with CORSIKA~\cite{CORSIKA} for the cosmic air shower simulation at the JUNO site, which gives the muon energy, momentum and multiplicity distributions arriving the surface.
Then MUSIC~\cite{KUDRYAVTSEV2009339} is employed to track muons traversing the rock to the underground experiment hall based on the local geological map.
The muon sample after transportation is used as the event generator of Geant4, with which the detector simulation is performed and all the secondary particles are recorded.
A simulation data set consisting of 16 million muon events is prepared, corresponding to about 50 days statistics.

\begin{table*}[htb]
\small
\begin{center}
	\begin{tabular}{c|c|c|c|c|c|c}
        \hline
    \multirow{2}{*}{Isotope} & \multirow{2}{*}{Decay mode}& \multirow{2}{*}{Decay energy~(MeV)} & \multirow{2}{*}{$\tau$} & \multicolumn{2}{c|}{Yield in LS~(/day)}  & \multirow{2}{*}{\minitab[c]{TFC fraction}} \\
    \cline{5-6}
    &&&&  Geant4 simulation & Scaled &  \\ \hline
    $^{12}$B & $\beta^-$       & 13.4  &  29.1~ms  & 1059  &   2282   & 90\%    \\ \hline
    $^{9}$Li & $\beta^-$: 50\% & 13.6  &  257.2~ms & 68    &   117    & 96\%    \\ \hline
    $^{9}$C  & $\beta^+$       & 16.5  &  182.5~ms & 21    &   160    & $>$99\% \\ \hline
    $^{8}$Li & $\beta^-+\alpha$& 16.0  &  1.21~s   & 725   &   649    & 94\%    \\ \hline
    $^{6}$He & $\beta^-$       & 3.5   &  1.16~s   & 526   &   2185   & 95\%    \\ \hline
    $^{8}$B  & $\beta^++\alpha$& $\sim$18 &  1.11~s& 35    &   447    & $>$99\% \\ \hline
    $^{10}$C & $\beta^+$       &3.6    &  27.8~s   & 816   &   878    & $>$99\% \\ \hline
    $^{11}$Be& $\beta^-$       & 11.5  &  19.9~s   & 9     &   59     & 96\%    \\ \hline
    $^{11}$C & $\beta^+$       & 1.98  &  29.4~min & 11811 & 46065    & 98\%    \\ \hline

	\end{tabular}
	\caption{Summary of the cosmogenic isotopes in JUNO. The isotope yields extracted from the Geant4 simulation, as well as the ones scaled to the measurements, are listed. The TFC fraction means the probability of finding at least one spallation neutron capture event between the muon and the isotope decay.}
	\label{tab:IsoYield}
\end{center}
\end{table*}

In the simulation the average muon track length in LS is about 23~m and the average deposited energy via ionization is 4.0~GeV.
Given the huge detector size, about 92\% of the 3.6~Hz LS muon events consist of one muon track, 6\% have two muon tracks, and the rest have more than two.
Events with more than one muon track are called muon bundles.
In general, muon tracks in one bundle are from the same air shower and are parallel.
In more than 85\% of the bundles, the distance between muon tracks is larger than 3~m.

\begin{figure}[htb]
\begin{center}
	\includegraphics[width=0.6\textwidth]{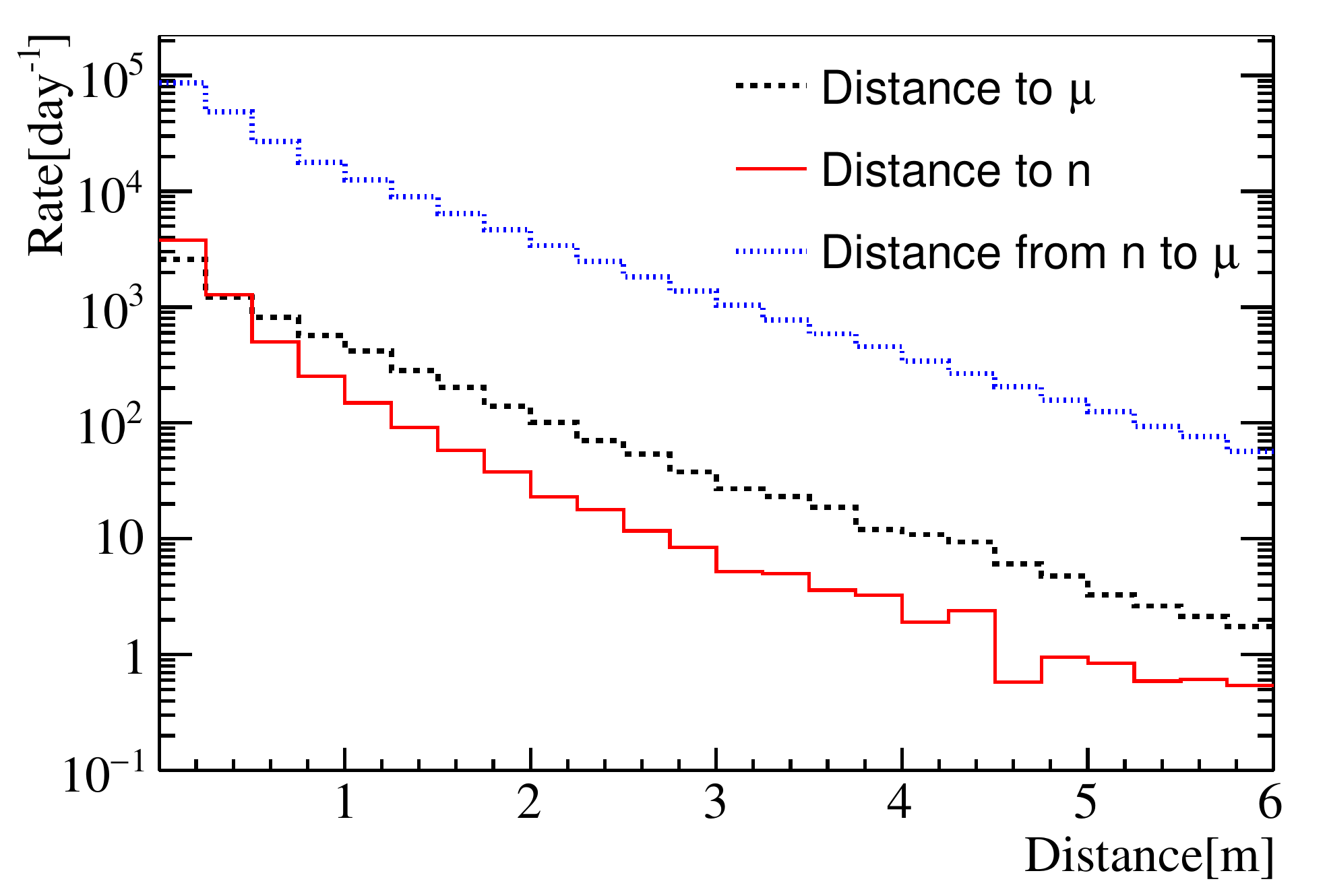}
	\caption{Distribution of the simulated distance between an isotope and its parent muon track~(black), and the distance between the isotope and the closest spallation neutron candidate~(red). The distance between a spallation neutron capture and its parent muon is also shown in blue. }\label{fig:IsotopeSpatial}
\end{center}
\end{figure}

The cosmogenic isotopes affecting this analysis are listed in Table~\ref{tab:IsoYield}.
The simulated isotope yields are found to be lower than those measured by KamLAND~\cite{Abe:2009aa} and Borexino~\cite{Bellini:2008mr}.
Thus, in our background estimation the yields are scaled to the results of the two experiments, by empirically modelling the production cross section as being proportional to $E_\mu^{0.74}$, where $E_\mu$ is the average energy of the muon at the detector.
Because the mean free paths of $\gamma$'s, $\pi$'s and neutrons are tens of centimeter in LS, the generation positions of the isotopes are close to the muon track, as shown in Fig.~\ref{fig:IsotopeSpatial}.
For more than 97\% of the isotopes, the distances are less than 3~m, leading to an effective cylindrical veto along the reconstructed muon track.
However, the veto time could only be set to 3 to 5~s to keep a reasonable detector live time, which removes a small fraction of $^{11}$C and $^{10}$C.
Thus, as mentioned before, the $^{11}$C, primarily from the $^{12}$C~($\gamma,~n$) reaction, with the largest yield and a long life time, will push the analysis threshold of the recoil electron to 2~MeV.
The removal of $^{10}$C, mainly generated in the $^{12}$C~($\pi^+,~np$) reaction, relies on the TFC among the muon, the neutron capture, and the isotope decay.

To perform the cylindrical volume veto, the muon track reconstruction is required.
There have been several reconstruction algorithms developed for JUNO as reported in Refs.~\cite{wonsak_bjorn_2018_1300496,Genster:2018caz,Zhang:2018kag,Wonsak:2018uby}.
A precision muon reconstruction algorithm was also developed in Double Chooz~\cite{ABE2014330}.
Based on these studies the muon reconstruction strategy in JUNO is assumed as:
1) If there is only one muon in the event, the track could be well reconstructed.
2) If there are two muons with a distance larger than 3~m in one event, which contributes 5.5\% to the total events, the two muons could be recognized and both well reconstructed.
If the distance is less than 3~m~(0.5\%), the number of muons could be identified via the energy deposit but only one track could be reconstructed.
3) If there are more than two muons in one event~(2\%), it is conservatively assumed no track information can be extracted and the whole detector would be vetoed for 1~s.
4) If the energy deposit is larger than 100~GeV~(0.1\%), no matter how many muons in the event, it is assumed no track can be reconstructed from such a big shower.

To design the TFC veto, the characteristics of neutron production are obtained from the simulation.
About 6\% of single muons and 18\% of muon bundles produce neutrons, and the average neutron numbers are about 11 and 15, respectively.
Most of the neutrons are close to each other, forming a neutron ball which can be used to estimate the shower position.
The simulated neutron yields are compared with the data in several experiments, such as Daya Bay~\cite{An:2017jng}, KamLAND~\cite{Abe:2009aa} and Borexino~\cite{Bellini:2013pxa}.
The differences are found to be less than 20\%.
The spatial distribution of the neutrons, defined as the distance between the neutron capture position and its parent muon track, are shown in Fig.~\ref{fig:IsotopeSpatial}.
More than 90\% of the neutrons are captured within 3~m from the muon track, consistent with KamLAND's measurement.
The advantage of LS detectors is the high detection efficiency of the neutron capture on hydrogen and carbon, which could be as high as 99\%.
If there is at least one neutron capture between the muon and the isotope decay, the event is defined as TFC tagged.
Then the TFC fraction is the ratio of the number of tagged isotopes to the total number of generated isotope.
In Table~\ref{tab:IsoYield} the TFC fraction in simulation is summarized.
The high TFC fraction comes from two aspects:
the first one is that a neutron and an isotope are simultaneously produced, like the $^{11}$C and $^{10}$C.
The other one is the coincidence between one isotope and the neutron(s) generated in the same shower.
If one isotope is produced, the medium number of neutrons generated by this muon is 13, and for more than one isotopes, the number of neutrons increases to 110, since isotopes are usually generated in showers with large energy deposit.
The red line in Fig.~\ref{fig:IsotopeSpatial} shows the distance between an isotope decay and the nearest neutron capture, which is mostly less than 2~m.

\subsubsection{Veto strategy}

Based on the information above, the muon veto strategy is designed as below.
\begin{itemize}
\item Whole detector veto:
	\subitem Veto 2~ms after every muon event, either passing through LS or water;
	\subitem Veto 1~s for the muon events without reconstructed tracks.
\item The cylindrical volume veto, depending on the distance~($d$) between the candidate and the muon track:
	\subitem Veto $d<$1~m for 5~s;
	\subitem Veto 1~m$<d<$3~m for 4~s;
	\subitem Veto 3~m$<d<$4~m for 2~s;
	\subitem Veto 4~m$<d<$5~m for 0.2~s.
\item The TFC veto:
	\subitem Veto a 2 meter spherical volume around a spallation neutron candidate for 160~s.
\end{itemize}
In the cylindrical volume veto, the 5~s and 4~s veto time is determined based on the life of \B~and $^{8}$Li.
The volume with $d$ between 4 and 5~m is mainly to remove $^{12}$B, which has a larger average distance because the primary generation process is $^{12}$C($n,p$) and neutrons have larger mean free path than $\gamma$'s and $\pi$'s.
For muon bundles with two muon tracks reconstructed, the above cylindrical volume veto will be applied to each track.
Compared to the veto strategies which reject any signal within a time window of 1.2~s and a 3~m cylinder along the muon track~\cite{Zhao:2016brs,An:2015jdp}, the above distance-dependent veto significantly improves the signal to background ratio.
The TFC veto is designed for the removal of $^{10}$C and $^{11}$Be.
Moreover, it effectively removes \B, $^{8}$Li, and $^{6}$He generated in large showers and muon bundles without track reconstruction abilities.

The muons not passing through LS, defined as external muons, will contribute to about 2\% of isotopes, concentrated at the edge of the LS.
Although there is no available muon track for the background suppression, the FV cut can effectively eliminate these isotopes and reach a background over signal ratio of less than 0.1\%, which can be safely neglected.

To estimate the dead time induced by the veto strategy and the residual background, a toy Monte Carlo sample is generated by mixing the \B~neutrino signal with the simulated muon data.
The whole detector veto and the cylindrical volume veto introduce 44\% dead time, while the TFC veto adds an additional 4\%.
The residual backgrounds above the 2~MeV analysis threshold consist of $^{12}$B, $^8$Li, $^6$He, $^{10}$C, and $^{11}$Be, as shown in Fig.~\ref{fig:CosmogenicIsotope}.
A potential improvement on the veto strategy may come from a joint likelihood based on the muon energy deposit density, the number of spallation neutrons, time and distance distributions between the isotope and muon, and among the isotope and neutrons.
In addition, this study could profit from the developing topological method for the discrimination between the signal~($e^-$) and background~($^{10}$C, $e^++\gamma$)~\cite{wonsak_bjorn_2018_1300496},

The actual isotope yields, distance distributions, and TFC fractions will be measured $in$-$situ$ in future.
Estimation of residual backgrounds and uncertainties will rely on these measurements.
Currently the systematic uncertainties are assumed based on KamLAND's measurements~\cite{Abe:2011em}, given the comparable overburden~(680~m and 1000~m): 1\% uncertainty to $^{12}$B, 3\% uncertainty to $^8$Li and $^6$He, and 10\% uncertainty to $^{10}$C and $^{11}$Be.

\subsection{Reactor antineutrinos}

The reactor antineutrino flux at JUNO site is about $2\times10^7$/cm$^2$/s assuming 36~GW thermal power.
Combining the oscillated antineutrino flux with the corresponding cross section~\cite{Vogel:1999zy}, the Inverse Beta Decay~(IBD) reaction rate between \nuebar~and proton is about 4~cpd/kt, and the elastic scattering rate between $\overline{\nu}_{x}$ and electron is about 1.9~cpd/kt in the energy range of 0 to 10~MeV.
The products of IBD reaction, $e^+$ and neutron, can be rejected to less than 0.5\% using the correlation between them.
The residual mainly comes from the two signals falling into one electronics readout window~(1~$\mu$s).
The recoil electron from the $\overline{\nu}-e$~ES channel, with a rate of 0.14~cpd/kt when the visible energy is larger than 2~MeV, cannot be distinguished from \B~$\nu$~signals.
A 2\% uncertainty is assigned to this background according to the uncertainties of antineutrino flux and the ES cross section.

\section{Expected results}
\label{Result}

\begin{figure}[htb]
\begin{center}
	\includegraphics[width=0.6\textwidth]{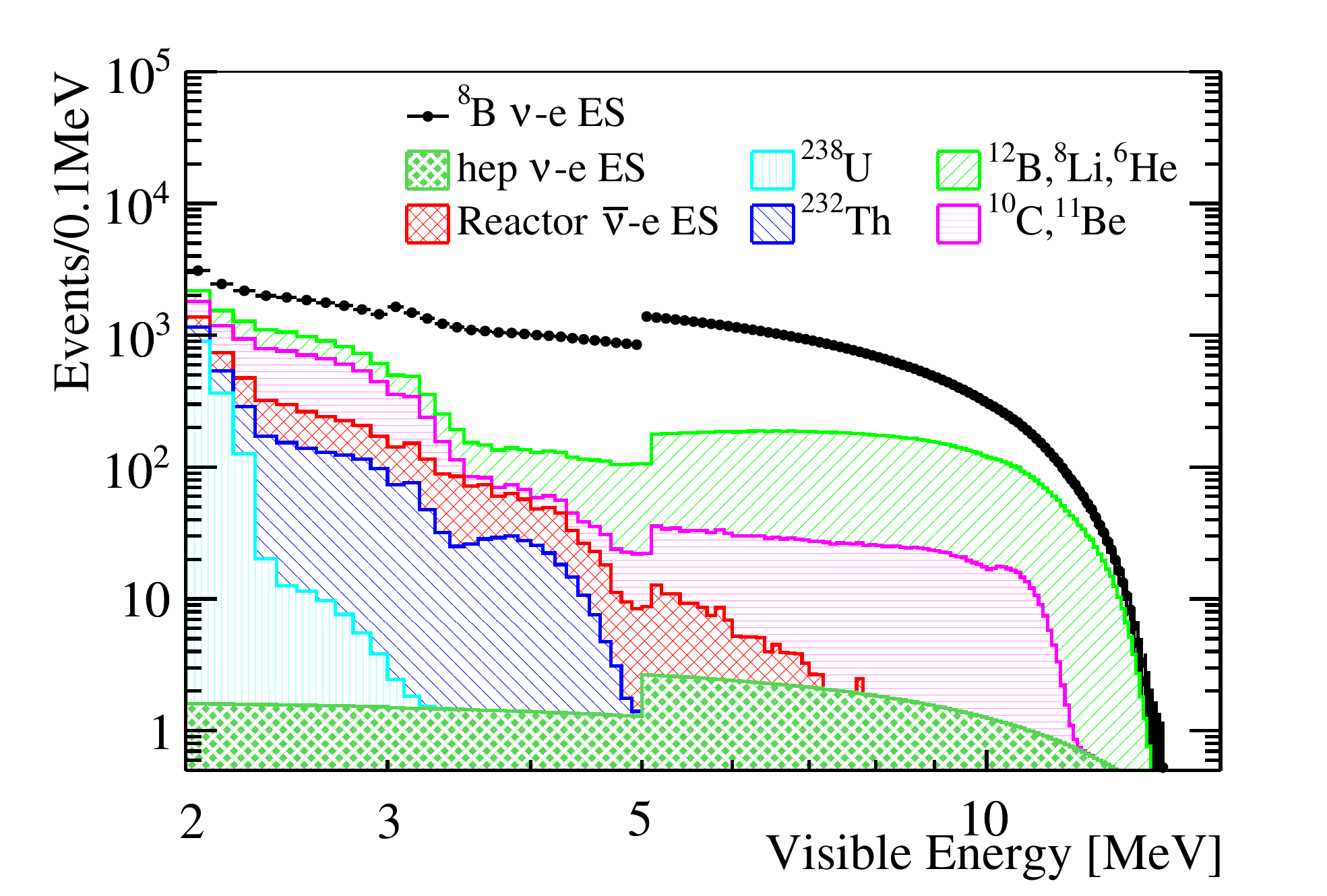}
	\caption{Expected signal and background spectra in ten years of data taking, with all selection cuts and muon veto methods applied. Signals are produced in the standard LMA-MSW framework using \dm=$4.8\times10^{-5}$~eV$^2$. The energy dependent fiducial volumes account for the discontinuities at 3~MeV and 5~MeV. }\label{fig:FinalSpec}
\end{center}
\end{figure}

After applying all the selection cuts, about 60,000 recoil electrons and 30,000 background events are expected in 10 years of data taking as listed in Table~\ref{tab:bkgSum} and shown in Fig.~\ref{fig:FinalSpec}.
The dead time due to muon veto is about 48\% in the whole energy range.
As listed in Table~\ref{tab:radioInternal}, the \BiL$-$\Tl~correlation cut removes 20\% of signals in the energy range of 3 to 5~MeV, and less than 2\% in other energy ranges.
The detection efficiency uncertainty, mainly from the FV cuts, is assumed to be 1\% according to Borexino's results\cite{Agostini:2017cav}.
Given that the uncertainty of the FV is determined using the uniformly distributed cosmogenic isotopes, the uncertainty is assumed to be correlated among the three energy-dependent FVs.
Since a spectrum distortion test will be performed, another important uncertainty source is the detector energy scale.
For electrons with energies larger than 2~MeV, the nonlinear relationship between the LS light output and the deposited energy is less than 1\%.
Moreover, electrons from the cosmogenic $^{12}$B decays, with an average energy of 6.4~MeV, can set strong constraints to the energy scale, as it was done in Daya Bay~\cite{Adey:2019zfo} and Double Chooz~\cite{DoubleChooz:2019qbj}.
Thus, a 0.3\% energy scale uncertainty is used in this analysis following the results in Ref.~\cite{Adey:2019zfo}.
Three analyses are reported based on these inputs.

\begin{table*}[tb]\small
  \begin{minipage}[c]{\textwidth}
  \resizebox{\textwidth}{!}{
	\begin{tabular}{c|c|c|ccccc|cc|c|c|cc}
        \hline
         \multirow{2}{*}{cpd/kt} & \multirow{2}{*}{FV} & \multirow{2}{*}{\B~signal eff.} &\multirow{2}{*}{$^{12}$B} & \multirow{2}{*}{$^{8}$Li} & \multirow{2}{*}{$^{10}$C} & \multirow{2}{*}{$^{6}$He} & \multirow{2}{*}{$^{11}$Be} & \multirow{2}{*}{$^{238}$U} & \multirow{2}{*}{$^{232}$Th} &  \multirow{2}{*}{$\overline{\nu}$-e ES} & \multirow{2}{*}{Total bkg.} & \multicolumn{2}{c}{Signal rate at }\\
           & & & &&&&&&&& & $\Delta m^{2\star}_{21}$ & $\Delta m^{2\dagger}_{21}$ \\ \hline
        (2,~3)~MeV & 7.9~kt& $\sim$51\%& 0.005 & 0.006 & 0.141 & 0.084 & 0.002 & 0.050 & 0.050 &  0.049 & 0.39 & 0.32 & 0.30 \\ \hline
        (3,~5)~MeV & 12.2~kt& $\sim$41\%&0.013 & 0.018 & 0.014 & 0.008 & 0.005 & 0 & 0.012 & 0.016 & 0.09 & 0.42 & 0.39 \\ \hline
        (5,~16)~MeV & 16.2~kt& $\sim$52\%& 0.065 & 0.085 & 0 & 0 & 0.023 & 0 & 0 & 0.002 & 0.17 & 0.61 & 0.59 \\ \hline
        Syst. error & 1\% & $<$1\% & 3\% & 10\% & 3\% & 10\% & 1\% & 1\% & 2\%  \\ \hline
	\end{tabular}}
	\caption{Summary of signal and background rates in different visible energy ranges with all selection cuts and muon veto methods applied. $\Delta m^{2\star}_{21}$= $4.8\times10^{-5}$~eV$^2$, and $\Delta m^{2\dagger}_{21}$= $7.5\times10^{-5}$~eV$^2$ }
	\label{tab:bkgSum}
\end{minipage}
\end{table*}

\subsection{Spectrum distortion test}

\begin{figure}[htb]
\begin{center}
	\includegraphics[width=0.6\textwidth]{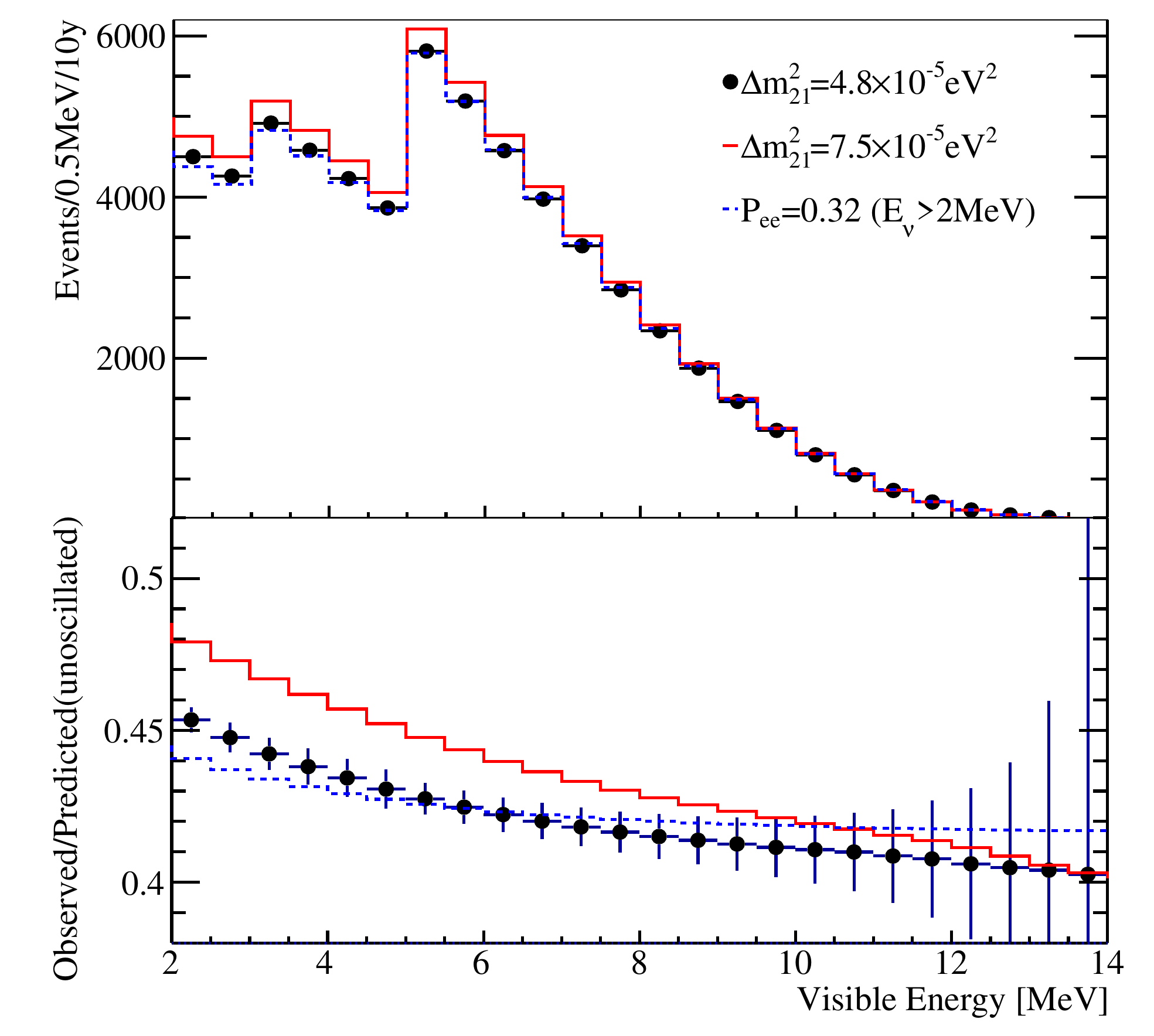}
	\caption{Background subtracted spectra produced in the standard LMA-MSW framework for two \dm~values~(black dots and red line, respectively), and the $P_{ee}=0.32~(E_\nu>2~{\rm MeV})$ assumption~(blue line). Their comparison with the no flavor conversion is shown in the bottom panel. Only statistical uncertainties are drawn. Details can be found in the text. }\label{fig:SpectrumTest}
\end{center}
\end{figure}

In the observed spectrum, the upturn comes from two aspects: the presence of $\nu_{\mu,\tau}$, and the upturn in $P_{ee}$.
A background-subtracted Asimov data set is produced in the standard LMA-MSW framework using \dm~=~$4.8\times10^{-5}$~eV$^2$, shown as the black points in the top panel of Fig.~\ref{fig:SpectrumTest}.
The other oscillation parameters are taken from PDG 2018~\cite{Tanabashi:2018oca}.
The error bars show only the statistical uncertainties.
The ratio to the prediction of no-oscillation is shown as the black points in the bottom panel of Fig.~\ref{fig:SpectrumTest}.
Here no-oscillation is defined as pure \nue~with an arrival flux of 5.25$\times10^6$/cm$^2$/s.
The signal rate variation with respect to the solar zenith angle has been averaged.
The expected signal spectrum using \dm~=~$7.5\times10^{-5}$~eV$^2$ is also drawn as the red line for comparison.
More signals can be found at the low energy range.
The spectral difference provides the sensitivity that enables measurement of \dm.

Since no upturn in the $P_{ee}$ was observed by previous experiments, a hypothesis test is performed.
The $P_{ee}$ is assumed as a flat value for neutrino energies larger than 2~MeV.
An example spectrum generated with $P_{ee}=0.32$ is drawn as the blue line in Fig.~\ref{fig:SpectrumTest}.
Comparing to the no-oscillation prediction, upturn of the blue line comes from the appearance of $\nu_{\mu,\tau}$'s, which have a different energy dependence in the $\nu-e$ ES cross section, as shown in Fig.~\ref{fig:Xsection}.
To quantify the sensitivity of rejecting this hypothesis, a $\chi^2$ statistic is constructed as:
\begin{equation}
\begin{split}
\chi^2= & 2\times\sum^{140}_{i=1}(N^i_{pre}-N^i_{obs}+N^i_{obs}\times \log\frac{N^i_{obs}}{N^i_{pre}})) +(\frac{\varepsilon_d}{\sigma_d})^2+(\frac{\varepsilon_f}{\sigma_f})^2+(\frac{\varepsilon_s}{\sigma_s})^2+(\frac{\varepsilon_e}{\sigma_e})^2+\sum^{10}_{j=1}(\frac{\varepsilon^j_b}{\sigma^j_b})^2 , \\
 N^i_{pre} &= (1+\varepsilon_d+\varepsilon_f+\alpha^i\times\varepsilon_s + \beta^i\times\frac{\varepsilon_e}{0.3\%})\times T_i
+ \sum^{10}_{j=1} (1+\varepsilon^j_b)\times B_{ij},
  \end{split}
\label{equ:spectrumT}
\end{equation}
where $N^i_{obs}$ is the observed number of events in the $i^{th}$ energy bin in the LMA-MSW framework, $N^i_{pre}$ is the predicted one in this energy bin, by adding the signal $T_i$ generated under the flat $P_{ee}$ hypothesis with the backgrounds $B_{ij}$, which is summing over $j$.
Systematic uncertainties are summarized in Table~\ref{table:syst}.
The detection efficiency uncertainty is $\sigma_d$~(1\%), the neutrino flux uncertainty is $\sigma_f$~(3.8\%), and $\sigma^j_b$ is the uncertainty of the $j^{th}$ background summarized in Table~\ref{tab:bkgSum}.
The corresponding nuisance parameters are $\varepsilon_d$, $\varepsilon_f$, and $\varepsilon^j_b$, respectively.

\begin{table}[h]
\begin{center}

	\begin{tabular}{c|c|c|c}
    \hline
	 & Notation & Value & Reference \\ \hline
	 Detection efficiency & $\sigma_d$ & 1\% & Borexino~\cite{Agostini:2017cav} \\
	 Detector energy scale & $\sigma_e$ & 0.3\% & Daya Bay~\cite{Adey:2019zfo}, Double Chooz~\cite{DoubleChooz:2019qbj} \\
	 The \B~$\nu$ flux & $\sigma_f$ & 3.8\% & SNO~\cite{Aharmim:2011vm} \\
	 The \B~$\nu$ spectrum shape & $\sigma_s$ & 1 & Ref.~\cite{Bahcall:1996qv} \\
	 The $j^{th}$ background & $\sigma^j_b$ & Table~\ref{tab:bkgSum} & This study \\
    \hline
    \end{tabular}
     \caption{ \label{table:syst} Summary of the systematic uncertainties. Since the uncertainty of \B~$\nu$ spectrum shape is absorbed in the coefficients $\alpha^i$, $\sigma_s$ equals to 1. See text for details.  }
\end{center}
\end{table}

The two uncertainties relating to the spectrum shape, the \B~$\nu$ spectrum shape uncertainty $\sigma_s$ and the detector energy scale uncertainty $\sigma_e$, are implemented in the statistic using coefficients $\alpha^i$ and $\beta^i$, respectively.
The neutrino energy spectrum with 1$\sigma$ deviation is converted to the visible spectrum of the recoil electron.
Its ratio to the visible spectrum converted from the nominal neutrino spectrum is denoted as $\alpha^i$.
In this way, the corresponding nuisance parameter, $\varepsilon_s$, follows the standard Gaussian distribution.
For the 0.3\% energy scale uncertainty, $\beta^i$ is derived from the ratio of the electron visible spectrum shifted by 0.3\% to the visible spectrum without shifting.

\begin{table}[h]
\begin{center}
	\begin{tabular}{c|c|c}
        \hline
    $\Delta \chi^2$ & $4.8\times10^{-5}$~eV$^2$&$7.5\times10^{-5}$~eV$^2$ \\
	\hline
    Stat.~only & 7.1  & 24.9 \\
    Stat.~+~\B~flux error &  6.8 &  24.2 \\
    Stat.~+~\B~shape error & 3.6 &  11.8 \\
	Stat.~+~energy scale error & 4.7 & 15.5 \\
    Stat.~+~background error  &  3.6 &  14.0 \\ \hline
    Final &  2.0 &  7.3 \\
    \hline
    \end{tabular}
     \caption{ \label{tab:spec} Rejection sensitivity for the flat $P_{ee}(E_\nu$$>$$2~{\rm MeV})$ hypothesis with 10 years data taking for the two \dm~values. The impact of each systematic uncertainty is also listed separately.  }
\end{center}
\end{table}

By minimizing the $\chi^2$, the abilities of excluding the flat $P_{ee}(E_\nu$$>$$2~{\rm MeV})$ hypothesis, in terms of $\Delta \chi^2$ values, are listed in Table~\ref{tab:spec}.
The total neutrino flux is constrained with a 3.8\% uncertainty~($\sigma_f$) from the SNO NC measurement, while the $P_{ee}$ value is free in the minimization.
The sensitivity is better at larger \dm~values due to the larger upturn in the visible energy spectrum.
If the true \dm~value is $7.5\times10^{-5}$~eV$^2$, the hypothesis could be rejected at 2.7$\sigma$ level.
%
%
The statistics-only sensitivity of rejecting the flat $P_{ee}$ hypothesis is $\Delta \chi^2= 4.9$~(18.9) for $\Delta m^2_{21}=4.8 \times 10^{-5}$~eV$^2$~($7.5 \times 10^{-5}$~eV$^2$) for 3~MeV threshold, comparing to $\Delta \chi^2= 7.1$~(24.9) for 2~MeV threshold.

To understand the effect of systematics, the impact of each systematic uncertainty is also provided in Table~\ref{tab:spec}.
The sensitivity is significantly reduced after introducing the systematics.
For instance, with \dm=$7.5\times10^{-5}$~eV$^2$, including the neutrino spectrum shape uncertainty~($\sigma_s$) almost halved the sensitivity, because the shape uncertainty could affect the ratio of events in the high and low visible energy ranges.
If the detector energy scale uncertainty~($\sigma_e$) is included, the sensitivity is also significantly reduced due to the same reason above.

\subsection{Day-Night asymmetry}

\begin{figure}[htb]
\begin{center}
	\includegraphics[width=0.6\textwidth]{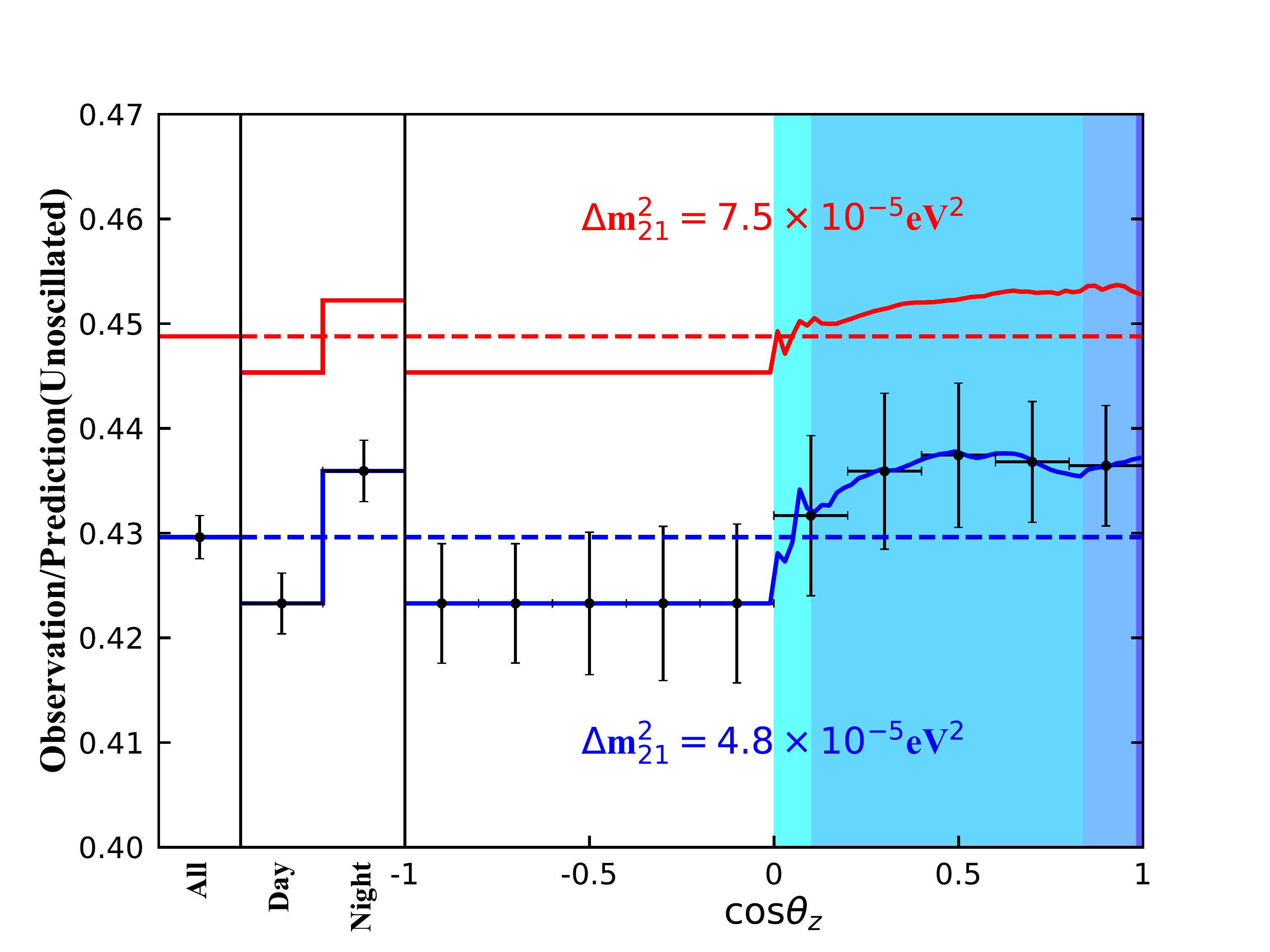}
	\caption{Ratio of \B~neutrino signals produced in the standard LMA-MSW framework to the no-oscillation prediction at different solar zenith angles. The uncertainties are propagated with a toy Monte Carlo simulation, and most of the systematic uncertainties are cancelled.}\label{fig:DayNightA}
\end{center}
\end{figure}

Solar neutrino propagation through the Earth is expected, via the MSW effect, to cause signal rate variation versus the solar zenith angle.
This rate variation observable also provides additional sensitivity to the \dm~value, as shown in Fig.~\ref{fig:DayNightA}.
The blue and red dashed lines represent the average ratio of the measured signal to the no-oscillation prediction, and they are calculated with \dm=$4.8\times10^{-5}$~eV$^2$ and $7.5\times10^{-5}$~eV$^2$, respectively.
The solid lines show the signal rate variations versus solar zenith angle.
Smaller \dm~values result in a larger MSW effect in the Earth and increased Day-Night asymmetry.
The error bars are the expected uncertainties.

The variation is quantified by defining the Day-Night asymmetry as:
\begin{equation}
A_{DN} = \frac{R_D-R_N}{(R_D+R_N)/2},
\label{equ:DNA}
\end{equation}
where $R_D$ and $R_N$ are the background-subtracted signal rates during the Day~($\cos\theta_z<0$) and Night~($\cos\theta_z>0$), respectively.
They are obtained by dividing the signal numbers listed in Table~\ref{table:DNA} by the effective exposure in Fig.~\ref{fig:zenithangle}.
The uncertainties are propagated with a toy Monte Carlo program to correctly include the correlation among systematics.
With ten years data taking, JUNO has the potential to observe the Day-Night asymmetry at a significance of 3$\sigma$ if \dm=$4.8\times10^{-5}$~eV$^2$.
Even when restricting to an energy range from 5 to 16~MeV, within which neither natural radioactivity nor $^{10}$C are significant, a 2.8$\sigma$ significance can still be achieved.
If \dm=$7.5\times10^{-5}$~eV$^2$, the expected $A_{DN}$ is $(-1.6\pm0.9)\%$ for the 2 to 16~MeV energy range.
The different $A_{DN}$ values also contribute to the \dm~determination.

The $A_{DN}$ measurement uncertainty is dominated by statistics since most of the systematic uncertainties are cancelled in the numerator and denominator.
Potential systematics could arise from different detector performance during Day and Night, however this is expected to be negligible for the LS detector.
Compared with Super-K's results from Ref.~\cite{Abe:2016nxk}, JUNO could reach the same precision of $A_{DN}$ in less than 10 years.
The primary improvement is a better signal over background ratio, because JUNO can reject \Tl~via the $\alpha$-$\beta$ cascade decay, and suppress cosmogenic isotopes via the TFC technique.

\begin{table}[h]
\begin{center}
	\begin{tabular}{c|c|c|c|c}
        \hline
		Energy & Exposure& Day & Night & $A_{DN}$ \\ \hline
		2$\sim$3~MeV & 41~kt$\cdot$y &4334 & 4428  & (-2.1 $\pm$3.2)\% \\
		3$\sim$5~MeV & 51~kt$\cdot$y &8686 & 8906  & (-2.5 $\pm$1.7)\% \\
		5$\sim$16~MeV& 84~kt$\cdot$y &17058& 17644  &(-3.4 $\pm$1.2)\% \\ \hline
		2$\sim$16~MeV& N/A& 30078& 30977 & (-2.9$\pm$0.9)\% \\ \hline
	\end{tabular}
	\caption{ Number of the background-subtracted signals during Day and Night in ten years of data taking for \dm~=~$4.8\times10^{-5}$~eV$^2$. A set of energy-dependent FV cuts are used and the values of the three energy ranges are provided. The uncertainties are dominated by signal and background statistics.}
	\label{table:DNA}
\end{center}
\end{table}

\subsection{Measurement of oscillation parameters}
\label{OscillationResults}
As mentioned above, in the standard neutrino oscillation framework, \dm~can be measured using the information in the spectra distortion and the signal rate variation versus solar zenith angle.
The signal rate versus visible energy and zenith angle~($\cos\theta_z$) is illustrated in Fig.~\ref{fig:2DSignal}.
To fit the distribution, a $\chi^2$ statistic is defined as:
\begin{equation}
\begin{split}
\chi^2 = &  2 \times\sum_{i=1}^{140}\sum_{j=1}^{100} \{ N^{pre}_{i,j} - N^{obs}_{i,j}
          +  N^{obs}_{i,j} \log \frac{ N^{obs}_{i,j}}{ N^{pre}_{i,j} } \} \\
		 &+ (\frac{\varepsilon_d}{\sigma_d})^2+(\frac{\varepsilon_f}{\sigma_f})^2+(\frac{\varepsilon_s}{\sigma_s})^2+(\frac{\varepsilon_e}{\sigma_e})^2+\sum^{10}_{k=1}(\frac{\varepsilon^k_b}{\sigma^k_b})^2 ,
\end{split}
\end{equation}
where $N^{pre}_{i,j}$ and $N^{obs}_{i,j}$ are the predicted and observed number of events in the $i^{th}$ energy bin and $j^{th}$~$\cos\theta_z$ bin, respectively.
The nuisance parameters have the same definitions as those in Eq.~\ref{equ:spectrumT}.
The oscillation parameters \T~and \dm~are obtained by minimizing the $\chi^2$.
The values of other oscillation parameters are from PDG 2018~\cite{Tanabashi:2018oca}, and their uncertainties are negligible in this study.

\begin{figure}[htb]
\vspace{0.8cm}
\begin{center}
	\includegraphics[width=0.6\textwidth]{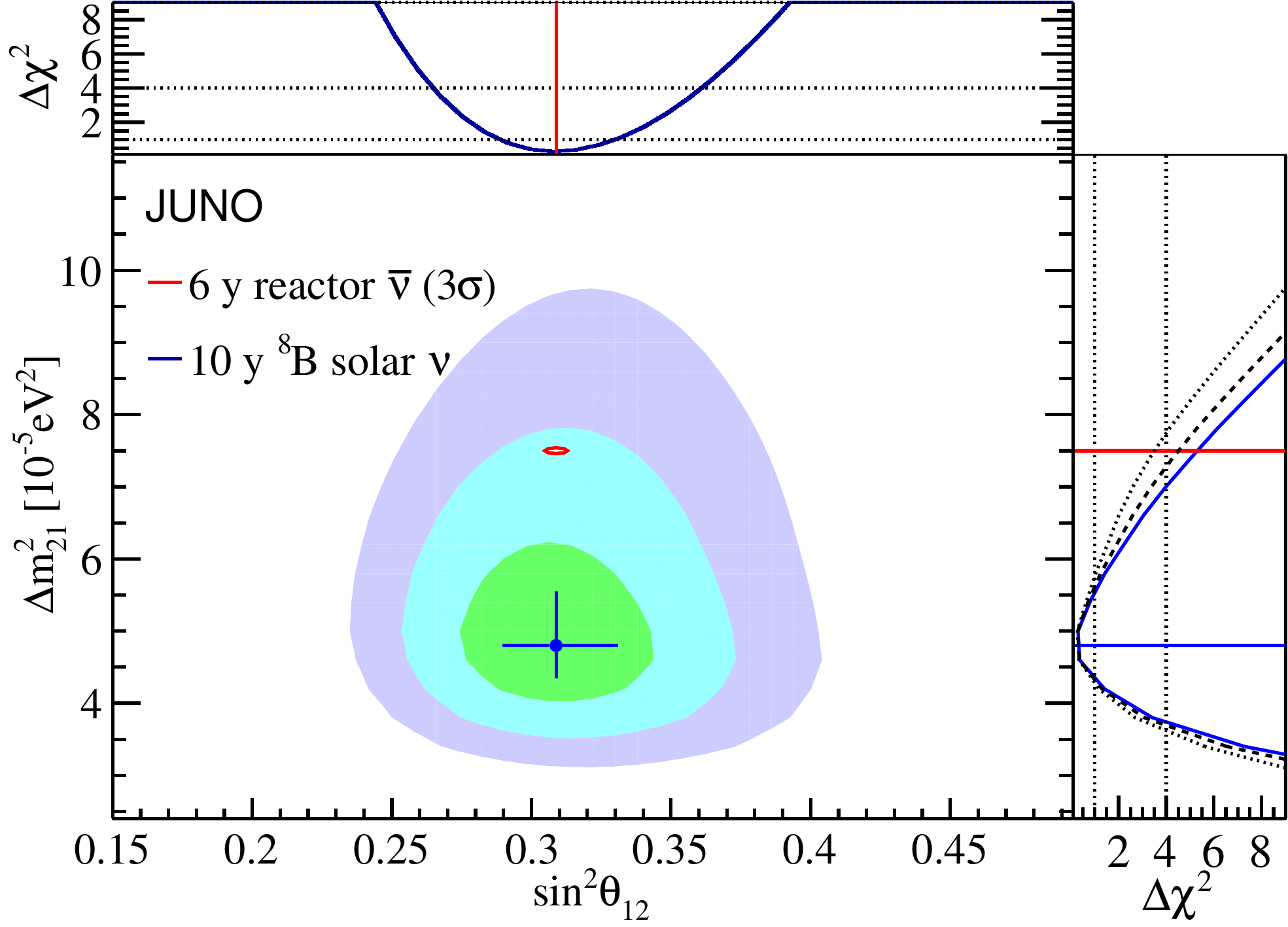}
	\caption{ 68.3\%, 95.5\%, and 99.7\% C.L.~allowed regions in the \T~and \dm~plane using the \B~solar neutrino in ten years data taking. The 99.7\% C.L.~region using six years reactor \nuebar~is drawn in red for comparison, in which the \dm~central value is set to the KamLAND's result~\cite{Gando:2013nba}, and uncertainties are taken from JUNO Yellow Book~\cite{An:2015jdp}. The one-dimensional $\Delta\chi^2$ for \T~and \dm~are shown in the top and right panels, respectively. The dashed line in the right panel represents the \dm~precision without \Tl~reduction, while the dotted line shows the results with an analysis threshold limited to 5~MeV due to intrinsic \U~and \Th~contaminations at 10$^{-15}$~g/g level. }\label{fig:Contour}
\end{center}
\vspace{0.8cm}
\end{figure}

With ten years of data taking the expected sensitivity of \T~and \dm~is shown in Fig.~\ref{fig:Contour}.
For \T, if the true value is 0.307, the 1$\sigma$ uncertainty is 0.023.
Since the sensitivity of \T~mainly comes from the comparison of the measured number of signals to the predicted one, about 60\% of its uncertainty is attributed to the \B~$\nu$  flux uncertainty $\sigma_f$.
For \dm, assuming a true value of $4.8\times10^{-5}$~eV$^2$ corresponds to a 68\% C.L.~region of~(4.3,~5.6)~$\times10^{-5}$~eV$^2$.
Assuming a true \dm~value of $7.5\times10^{-5}$~eV$^2$ corresponds to a 68\% C.L.~region of (6.3,~9.1)~$\times10^{-5}$~eV$^2$.
The asymmetric uncertainty arises because the Day-Night asymmetry measurement plays a more important role with a smaller \dm.
The \dm~precision is mainly limited by the statistical uncertainty on the Day-Night asymmetry measurement,
with the signal statistics responsible for about 50\% of the uncertainty.
The subdominant uncertainty of 25\% arises from the \B~$\nu$ flux uncertainty, with about a 10\% contribution from the \B~$\nu$ spectrum shape uncertainty.
In conclusion, the discrimination sensitivity between the above two \dm~values reaches more than 2$\sigma$~($\Delta\chi^2\sim5.3$), similar to the current solar global fitting results~\cite{Esteban:2018azc}.

A crucial input to this study is the LS intrinsic radioactivity level.
The current result is based on the assumption of achieving $10^{-17}$~g/g \U~and \Th, and a 2,600~cpd/kt $^{210}$Po decay rate.
If the $^{210}$Po decay rate reaches to more than 10,000~cpd/kt like the Phase I of Borexino, \Tl~could not be reduced by the \BiL$-$\Tl~cascade decay, and the S/B ratio decreases from 35 to 0.6 in the 3 to 5~MeV energy range.
The effect on the \dm~precision is shown as the dashed line in the right panel of Fig.~\ref{fig:Contour}.
If the \U~and \Th~contaminations are at $10^{-15}$~g/g level, the analysis threshold would be limited to 5~MeV.
The \dm~measurement would mainly rely on the Day-Night asymmetry, and the projected sensitivity is shown as the dotted line in the right panel of Fig.~\ref{fig:Contour}.
In this case the sensitivity of distinguishing the two \dm~values is slightly worse than 2$\sigma$~($\Delta\chi^2\sim 3.5$).
However, if \U~and \Th~contaminations are smaller than 10$^{-17}$~g/g, the sensitivities do not have significantly improvements.
Because the background in the 2 to 5~MeV energy range is dominated by cosmogenic $^{10}$C and $^{6}$He in this case.

\section{Summary and prospects}

More than fifty years after the discovery of solar neutrinos, they continue to provide the potential for major contributions to neutrino physics.
The JUNO experiment, with a 20~kt LS detector, can shed light on the current tension between \dm~values measured using solar neutrinos and reactor antineutrinos.
Compared to the discussion in JUNO Yellow Book~\cite{An:2015jdp}, a set of energy-dependent FV cuts is newly designed based on comprehensive background studies, leading to the maximized target mass with negligible external background.
The veto strategies for cosmogenic isotopes are also improved compared to those in Refs.~\cite{Zhao:2016brs,An:2015jdp}.
A set of distance-dependent veto time cuts are developed for the cylindrical veto along the muon track, resulting in a significantly improved signal to background ratio.
With 10$^{-17}$~g/g intrinsic \U~and \Th, the analysis threshold of recoil electrons from the ES channel can be lowered from the current 3~MeV in Borexino~\cite{Agostini:2017cav} to 2~MeV.
In the standard three-flavor neutrino oscillation framework, the spectrum distortion and the Day-Night asymmetry lead to a \dm~measurement of $4.8^{+0.8}_{-0.5}~(7.5^{+1.6}_{-1.2})\times10^{-5}$~eV$^2$, with a similar precision to the current solar global fitting result.

The interactions between neutrinos and carbon, such as $\nu_x$-$^{12}$C NC and \nue-$^{13}$C CC channels, are under investigation.
Most of the neutrino energy is carried by electrons in the CC reactions, and can also be used for the spectrum distortion examination.
Furthermore, both channels could be utilized in the search for $hep$ solar neutrinos, which has a predicted arrival flux of $8.25\times10^3$~/cm$^2$/s but has not been detected yet.

\section*{Acknowledgement}
\addcontentsline{toc}{section}{Acknowledgement}

We are grateful for the ongoing cooperation from the China General Nuclear Power Group.
This work was supported by
the Chinese Academy of Sciences,
the National Key R\&D Program of China,
the CAS Center for Excellence in Particle Physics,
the Joint Large-Scale Scientific Facility Funds of the NSFC and CAS,
Wuyi University,
and the Tsung-Dao Lee Institute of Shanghai Jiao Tong University in China,
the Institut National de Physique Nucl\'eaire et de Physique de Particules (IN2P3) in France,
the Istituto Nazionale di Fisica Nucleare (INFN) in Italy,
the Fond de la Recherche Scientifique (F.R.S-FNRS) and FWO under the ``Excellence of Science ¨C EOS¡± in Belgium,
the Conselho Nacional de Desenvolvimento Cient\'ifico e Tecnol\`ogico in Brazil,
the Agencia Nacional de Investigaci\'on y Desarrollo in Chile,
the Charles University Research Centre and the Ministry of Education, Youth, and Sports in Czech Republic,
the Deutsche Forschungsgemeinschaft (DFG), the Helmholtz Association, and the Cluster of Excellence PRISMA+ in Germany,
the Joint Institute of Nuclear Research (JINR), Lomonosov Moscow State University, and Russian Foundation for Basic Research (RFBR) in Russia,
the MOST and MOE in Taiwan,
the Chulalongkorn University and Suranaree University of Technology in Thailand,
and the University of California at Irvine in USA.

\bibliographystyle{unsrt}

\bibliography{JUNOB8Solar}

\end{document}

%% file: author.tex
\author[5]{Angel Abusleme}
\author[45]{Thomas Adam}
\author[67]{Shakeel Ahmad}
\author[55]{Sebastiano Aiello}
\author[67]{Muhammad Akram}
\author[67]{Nawab Ali}
\author[29]{Fengpeng An}
\author[10]{Guangpeng An}
\author[22]{Qi An}
\author[55]{Giuseppe Andronico}
% ...
\affil[5]{Pontificia Universidad Cat\'{o}lica de Chile, Santiago, Chile}
% ...
\affil[10]{Institute of High Energy Physics, Beijing, China}
% ...
\affil[22]{University of Science and Technology of China, Hefei, China}
% ...
\affil[29]{East China University of Science and Technology, Shanghai, China}
% ...
\affil[45]{IPHC, Universit\'{e} de Strasbourg, CNRS/IN2P3, F-67037 Strasbourg, France}
% ...
\affil[55]{INFN Catania and Dipartimento di Fisica e Astronomia dell Universit\`{a} di Catania, Catania, Italy}
% ...
\affil[67]{Pakistan Institute of Nuclear Science and Technology, Islamabad, Pakistan}
% ...


%% file: JUNOB8Solar.bbl
\begin{thebibliography}{10}

\bibitem{Davis:1968cp}
Raymond Davis, Don~S. Harmer, and Kenneth~C. Hoffman.
\newblock Search for neutrinos from the sun.
\newblock {\em Phys. Rev. Lett.}, 20:1205--1209, May 1968.

\bibitem{Hirata:1989zj}
K.~S. Hirata et~al.
\newblock {Observation of B-8 Solar Neutrinos in the Kamiokande-II Detector}.
\newblock {\em Phys. Rev. Lett.}, 63:16, 1989.

\bibitem{Anselmann:1992um}
P.~Anselmann et~al.
\newblock {Solar neutrinos observed by GALLEX at Gran Sasso.}
\newblock {\em Phys. Lett. B}, 285:376--389, 1992.

\bibitem{Altmann:2000ft}
M.~Altmann et~al.
\newblock {GNO solar neutrino observations: Results for GNO I}.
\newblock {\em Phys. Lett. B}, 490:16--26, 2000.

\bibitem{Abazov:1991rx}
A.~I. Abazov et~al.
\newblock {Search for neutrinos from sun using the reaction Ga-71
  (electron-neutrino e-) Ge-71}.
\newblock {\em Phys. Rev. Lett.}, 67:3332--3335, 1991.

\bibitem{Fukuda:1998fd}
Y.~Fukuda et~al.
\newblock {Measurements of the solar neutrino flux from Super-Kamiokande's
  first 300 days}.
\newblock {\em Phys. Rev. Lett.}, 81:1158--1162, 1998.

\bibitem{Ahmad:2001an}
Q.~R. Ahmad et~al.
\newblock {Measurement of the rate of $\nu_e+d \to p+p+e^-$ interactions
  produced by $^8B$ solar neutrinos at the Sudbury Neutrino Observatory}.
\newblock {\em Phys. Rev. Lett.}, 87:071301, 2001.

\bibitem{Ahmad:2002jz}
Q.~R. Ahmad et~al.
\newblock {Direct evidence for neutrino flavor transformation from neutral
  current interactions in the Sudbury Neutrino Observatory}.
\newblock {\em Phys. Rev. Lett.}, 89:011301, 2002.

\bibitem{Arpesella:2007xf}
C.~Arpesella et~al.
\newblock {First real time detection of Be-7 solar neutrinos by Borexino}.
\newblock {\em Phys. Lett. B}, 658:101--108, 2008.

\bibitem{Chen:1985na}
H.~H. Chen.
\newblock {Direct Approach to Resolve the Solar Neutrino Problem}.
\newblock {\em Phys. Rev. Lett.}, 55:1534--1536, 1985.

\bibitem{Villante:2013mba}
Francesco~L. Villante, Aldo~M. Serenelli, Franck Delahaye, and Marc~H.
  Pinsonneault.
\newblock {The chemical composition of the Sun from helioseismic and solar
  neutrino data}.
\newblock {\em Astrophys. J.}, 787:13, 2014.

\bibitem{Bergemann2014}
Maria Bergemann and Aldo Serenelli.
\newblock {\em Solar Abundance Problem}, pages 245--258.
\newblock Springer International Publishing, Cham, 2014.

\bibitem{Wolfenstein:1978ue}
L.~Wolfenstein.
\newblock Neutrino oscillations in matter.
\newblock {\em Phys. Rev. D}, 17:2369--2374, May 1978.

\bibitem{Mikheev:1986gs}
S.~P. Mikheyev and A.~{\relax Yu}. Smirnov.
\newblock {Resonance Amplification of Oscillations in Matter and Spectroscopy
  of Solar Neutrinos}.
\newblock {\em Sov. J. Nucl. Phys.}, 42:913--917, 1985.

\bibitem{Maltoni:2015kca}
Michele Maltoni and Alexei~Yu. Smirnov.
\newblock {Solar neutrinos and neutrino physics}.
\newblock {\em Eur. Phys. J. A}, 52(4):87, 2016.

\bibitem{Aharmim:2011vm}
B.~Aharmim et~al.
\newblock {Combined Analysis of all Three Phases of Solar Neutrino Data from
  the Sudbury Neutrino Observatory}.
\newblock {\em Phys. Rev. C}, 88:025501, 2013.

\bibitem{Abe:2016nxk}
K.~Abe et~al.
\newblock {Solar Neutrino Measurements in Super-Kamiokande-IV}.
\newblock {\em Phys. Rev. D}, 94(5):052010, 2016.

\bibitem{Gando:2013nba}
A.~Gando et~al.
\newblock {Reactor On-Off Antineutrino Measurement with KamLAND}.
\newblock {\em Phys. Rev. D}, 88(3):033001, 2013.

\bibitem{An:2015jdp}
Fengpeng An et~al.
\newblock {Neutrino Physics with JUNO}.
\newblock {\em J. Phys. G}, 43(3):030401, 2016.

\bibitem{Agostini:2017cav}
M.~Agostini et~al.
\newblock {Improved measurement of $^8$B solar neutrinos with
  1.5  kt$\cdot$y of Borexino exposure}.
\newblock {\em Phys. Rev. D}, 101(6):062001, 2020.

\bibitem{Esteban:2018azc}
Ivan Esteban, M.~C. Gonzalez-Garcia, Alvaro Hernandez-Cabezudo, Michele
  Maltoni, and Thomas Schwetz.
\newblock {Global analysis of three-flavour neutrino oscillations: synergies
  and tensions in the determination of $\theta_{23}$, $\delta_{CP}$, and the
  mass ordering}.
\newblock {\em JHEP}, 01:106, 2019.

\bibitem{Bahcall:1996qv}
John~N. Bahcall, E.~Lisi, D.~E. Alburger, L.~De~Braeckeleer, S.~J. Freedman,
  and J.~Napolitano.
\newblock {Standard neutrino spectrum from B-8 decay}.
\newblock {\em Phys. Rev. C}, 54:411--422, 1996.

\bibitem{Bahcall:1997eg}
John~N. Bahcall.
\newblock {Gallium solar neutrino experiments: Absorption cross-sections,
  neutrino spectra, and predicted event rates}.
\newblock {\em Phys. Rev. C}, 56:3391--3409, 1997.

\bibitem{Bahcall:2004pz}
John~N. Bahcall, Aldo~M. Serenelli, and Sarbani Basu.
\newblock {New solar opacities, abundances, helioseismology, and neutrino
  fluxes}.
\newblock {\em Astrophys. J.}, 621:L85--L88, 2005.

\bibitem{PyEphem}
Brandon R.
\newblock Pyephem astronomy library.

\bibitem{aacgmv2}
S.~G. Shepherd.
\newblock Altitude-adjusted corrected geomagnetic coordinates: Definition and
  functional approximations.
\newblock {\em Journal of Geophysical Research: Space Physics},
  119(9):7501--7521, 2014.

\bibitem{Ioannisian:2015qwa}
A.N. Ioannisian, A.~Yu. Smirnov, and D.~Wyler.
\newblock {Oscillations of the $^7$Be solar neutrinos inside the Earth}.
\newblock {\em Phys. Rev. D}, 92(1):013014, 2015.

\bibitem{Dziewonski:1981xy}
A.~M. Dziewonski and D.~L. Anderson.
\newblock {Preliminary reference earth model}.
\newblock {\em Phys. Earth Planet. Interiors}, 25:297--356, 1981.

\bibitem{Tanabashi:2018oca}
M.~Tanabashi et~al.
\newblock {Review of Particle Physics}.
\newblock {\em Phys. Rev. D}, 98(3):030001, 2018.

\bibitem{Giunti:2007ry}
Carlo Giunti and Chung~W. Kim.
\newblock {\em {Fundamentals of Neutrino Physics and Astrophysics}}.
\newblock 4 2007.

\bibitem{Adey:2019zfo}
D.~Adey et~al.
\newblock {A high precision calibration of the nonlinear energy response at
  Daya Bay}.
\newblock {\em Nucl. Instrum. Meth. A}, 940:230--242, 2019.

\bibitem{sisti_monica_2018_1300598}
Monica Sisti.
\newblock {Radioactive background control for the JUNO experimental setup},
  June 2018.

\bibitem{Geant4}
S.~Agostinelli et~al.
\newblock {GEANT4: A Simulation toolkit}.
\newblock {\em Nucl. Instrum. Meth.}, A506:250--303, 2003.

\bibitem{Li_2016}
Xin-Ying Li, Zi-Yan Deng, Liang-Jian Wen, Wei-Dong Li, Zheng-Yun You, Chun-Xu
  Yu, Yu-Mei Zhang, and Tao Lin.
\newblock Simulation of natural radioactivity backgrounds in the {JUNO} central
  detector.
\newblock {\em Chinese Physics C}, 40(2):026001, February 2016.

\bibitem{Zhang:2017ocm}
Xuantong Zhang, Jie Zhao, Shulin Liu, Shunli Niu, Xiaoming Han, Liangjian Wen,
  Jincheng He, and Tao Hu.
\newblock {Study on the large area MCP-PMT glass radioactivity reduction}.
\newblock {\em Nucl. Instrum. Meth. A}, 898:67--71, 2018.

\bibitem{Abe:2011em}
S.~Abe et~al.
\newblock {Measurement of the $^8$B Solar Neutrino Flux with the KamLAND Liquid
  Scintillator Detector}.
\newblock {\em Phys. Rev. C}, 84:035804, 2011.

\bibitem{Zhao:2013mba}
Jie Zhao, Ze-Yuan Yu, Jiang-Lai Liu, Xiao-Bo Li, Fei-Hong Zhang, and Dong-Mei
  Xia.
\newblock {$^{13}$C($\alpha$,n)$^{16}$O background in a liquid scintillator
  based neutrino experiment}.
\newblock {\em Chin. Phys. C}, 38(11):116201, 2014.

\bibitem{Distillation}
P.~Lombardi et~al.
\newblock {Distillation and stripping pilot plants for the JUNO neutrino
  detector: Design, operations and reliability}.
\newblock {\em Nucl. Instrum. Meth.}, A925:6--17, 2019.

\bibitem{OSIRIS}
Alex G$\ddot{\rm o}$ttel et~al.
\newblock A 20 ton liquid scintillator detector as a radioactivity monitor for
  juno.

\bibitem{Arpesella:2008mt}
C.~Arpesella et~al.
\newblock {Direct Measurement of the Be-7 Solar Neutrino Flux with 192 Days of
  Borexino Data}.
\newblock {\em Phys. Rev. Lett.}, 101:091302, 2008.

\bibitem{Agostini:2017ixy}
M.~Agostini et~al.
\newblock {First Simultaneous Precision Spectroscopy of $pp$, $^7$Be, and $pep$
  Solar Neutrinos with Borexino Phase-II}.
\newblock {\em Phys. Rev. D}, 100(8):082004, 2019.

\bibitem{Bellato:2020lio}
M.~Bellato et~al.
\newblock {Embedded Readout Electronics R\&D for the Large PMTs in the JUNO
  Experiment}.
\newblock March 2020.

\bibitem{ATOM}
Table of nuclides.

\bibitem{Balata:2006ue}
H.~Back et~al.
\newblock {CNO and pep neutrino spectroscopy in Borexino: Measurement of the
  deep underground production of cosmogenic $^{11}$C in organic liquid
  scintillator}.
\newblock {\em Phys. Rev. C}, 74:045805, 2006.

\bibitem{Agostini:2018uly}
M.~Agostini et~al.
\newblock {Comprehensive measurement of $pp$-chain solar neutrinos}.
\newblock {\em Nature}, 562(7728):505--510, 2018.

\bibitem{Abe:2009aa}
S.~Abe et~al.
\newblock {Production of Radioactive Isotopes through Cosmic Muon Spallation in
  KamLAND}.
\newblock {\em Phys. Rev. C}, 81:025807, 2010.

\bibitem{Bellini:2008mr}
G.~Bellini et~al.
\newblock {Measurement of the solar $^8$B neutrino rate with a liquid
  scintillator target and 3 MeV energy threshold in the Borexino detector}.
\newblock {\em Phys. Rev. D}, 82:033006, 2010.

\bibitem{Li:2015kpa}
Shirley~Weishi Li and John~F. Beacom.
\newblock {Spallation Backgrounds in Super-Kamiokande Are Made in Muon-Induced
  Showers}.
\newblock {\em Phys. Rev. D}, 91(10):105005, 2015.

\bibitem{Li:2015lxa}
Shirley~Weishi Li and John~F. Beacom.
\newblock {Tagging Spallation Backgrounds with Showers in Water-Cherenkov
  Detectors}.
\newblock {\em Phys. Rev. D}, 92(10):105033, 2015.

\bibitem{CORSIKA}
D.~{Heck}, J.~{Knapp}, J.~N. {Capdevielle}, G.~{Schatz}, and T.~{Thouw}.
\newblock {\em {CORSIKA: a Monte Carlo code to simulate extensive air
  showers.}}
\newblock 1998.

\bibitem{KUDRYAVTSEV2009339}
V.A. Kudryavtsev.
\newblock Muon simulation codes music and musun for underground physics.
\newblock {\em Computer Physics Communications}, 180(3):339 -- 346, 2009.

\bibitem{wonsak_bjorn_2018_1300496}
Bj$\ddot{\rm o}$rn Wonsak.
\newblock {3D Topological Reconstruction for the JUNO Detector}, June 2018.

\bibitem{Genster:2018caz}
Christoph Genster, Michaela Schever, Livia Ludhova, Michael Soiron, Achim
  Stahl, and Christopher Wiebusch.
\newblock {Muon reconstruction with a geometrical model in JUNO}.
\newblock {\em JINST}, 13(03):T03003, 2018.

\bibitem{Zhang:2018kag}
Kun Zhang, Miao He, Weidong Li, and Jilei Xu.
\newblock {Muon Tracking with the fastest light in the JUNO Central Detector}.
\newblock {\em Radiation Detection Technology and Methods}, 2(13), March 2018.

\bibitem{Wonsak:2018uby}
Bj$\ddot{\rm o}$rn~S. Wonsak, Caren~I. Hagner, Dominikus~A. Hellgartner, Kai
  Loo, Sebastian Lorenz, David~J. Meyh$\ddot{\rm o}$fer, Lothar Oberauer,
  Henning Rebber, Wladyslaw~H. Trzaska, and Michael Wurm.
\newblock {Topological track reconstruction in unsegmented, large-volume liquid
  scintillator detectors}.
\newblock {\em JINST}, 13(07):P07005, 2018.

\bibitem{ABE2014330}
Y.~Abe et~al.
\newblock {Precision Muon Reconstruction in Double Chooz}.
\newblock {\em Nucl. Instrum. Meth. A}, 764:330--339, 2014.

\bibitem{An:2017jng}
Fengpeng An et~al.
\newblock {Cosmogenic neutron production at Daya Bay}.
\newblock {\em Phys. Rev. D}, 97(5):052009, 2018.

\bibitem{Bellini:2013pxa}
G.~Bellini et~al.
\newblock {Cosmogenic Backgrounds in Borexino at 3800 m water-equivalent
  depth}.
\newblock {\em JCAP}, 1308:049, 2013.

\bibitem{Zhao:2016brs}
Jie Zhao, Liang-Jian Wen, Yi-Fang Wang, and Jun Cao.
\newblock {Physics potential of searching for $0\nu\beta\beta$ decays in JUNO}.
\newblock {\em Chin. Phys. C}, 41(5):053001, 2017.

\bibitem{Vogel:1999zy}
P.~Vogel and John~F. Beacom.
\newblock {Angular distribution of neutron inverse beta decay, anti-neutrino(e)
  + p $\rightarrow$ e$^+$ + n}.
\newblock {\em Phys. Rev. D}, 60:053003, 1999.

\bibitem{DoubleChooz:2019qbj}
H.~de~Kerret et~al.
\newblock {First Double Chooz $\mathbf{\theta_{13}}$ Measurement via Total
  Neutron Capture Detection}.
\newblock {\em Nature Phys.}, 16(5):558--564, 2020.

\end{thebibliography}
